\documentclass[12pt]{article}

\pdfoutput=1

\usepackage[top=80pt,bottom=85pt,left=85pt,right=85pt]{geometry}
\usepackage{amsmath,amssymb,graphicx,float,color,tikz,cite,savesym,wasysym}
\usepackage{caption}
\usepackage{subcaption}
\usepackage[debug,pageanchor=false]{hyperref}
\definecolor{link}{rgb}{.8,.15,.1}
\hypersetup{colorlinks=true,linkcolor=link,citecolor=link,urlcolor=link,linktocpage}

\newcommand{\ii}{\mathrm{i}}
\newcommand{\dd}{\mathrm{d}}

\makeatletter
\@addtoreset{equation}{section}
\makeatother

\begin{document}

	\begin{titlepage}

	\begin{center}

	\vskip .5in 
	\noindent

	{\Large \bf{Higher spins and Finsler geometry}}

	\bigskip\medskip

	 Alessandro Tomasiello\\

	\bigskip\medskip
	{\small 
Dipartimento di Matematica, Universit\`a degli Studi di Milano--Bicocca, \\ Via Cozzi 55, 20126 Milano, Italy \\ and \\ INFN, sezione di Milano--Bicocca
		}

	\vskip .5cm 
	{\small \tt alessandro.tomasiello@unimib.it}
	\vskip .9cm 
	\end{center}

	\noindent

Finsler geometry is a natural generalization of (pseudo-)Riemannian geometry, where the line element is not the square root of a quadratic form but a more general homogeneous function. Parameterizing this in terms of symmetric tensors suggests a possible interpretation in terms of higher-spin fields. We will see here that, at linear level in these fields, the Finsler version of the Ricci tensor leads to the curved-space Fronsdal equation for all spins, plus a Stueckelberg-like coupling. Nonlinear terms can also be systematically analyzed, suggesting a possible interacting structure. No particular choice of spacetime dimension is needed. The Stueckelberg mechanism breaks gauge transformations to a redundancy that does not change the geometry. This creates a serious issue: non-transverse modes are not eliminated, at least for the versions of Finsler dynamics examined in this paper.

	\noindent

	\vfill
	\eject

	\end{titlepage}

\tableofcontents

\section{Introduction} 
\label{sec:intro}

The dynamics of fields with spin higher than two has an intricate structure. Their free equations of motion are already quite complicated when they have a mass \cite{hagen-singh}, but simplifies in the massless case, where their equation of motion mimics that of spin two, using the so-called Fronsdal operator ${\mathcal F}$ \cite{fronsdal}. Their interactions have a highly constrained dynamics; for reviews see for example \cite{sagnotti-notes-hs,sorokin-hs,ponomarev-hs,bekaert-cnockaert-iazeolla-vasiliev, giombi-tasi}. It has long been surmised that they are important for theories of quantum gravity. String theory does have a tower of such fields, with masses related to the string tension. On the other hand, theories of Vasiliev type describe massless higher spins in (A)dS \cite{fradkin-vasiliev1,fradkin-vasiliev2,vasiliev1,vasiliev2,vasiliev3}. Their status as quantum theories is bolstered by their holographic interpretation \cite{klebanov-polyakov,giombi-yin}, although some degree of non-locality emerges \cite{boulanger-kessel-skvortsov-taronna,sleight-taronna-locality}. Higher spin fields and symmetries are expected more generally in holography, although most often in a broken version \cite{maldacena-zhiboedov}. Some versions of these theories can indeed be even obtained as limits of string theory \cite{giombi-minwalla-prakash-trivedi-wadia-yin,chang-minwalla-sharma-yin}. Restrictions to local covariant massless higher spins in flat space were found in \cite{bekaert-boulanger-leclercq,roiban-tseytlin,ponomarev-selfdual}; more recently, progress has been made using light-cone approach \cite{ponomarev-skvortsov,ponomarev-selfdual}.

Finsler geometry is a natural generalization of Riemannian and pseudo-Riemannian geometry; for some recent introductions see \cite{chern-shen,bao-chern-shen,shen-shen}. In the former case, instead of writing the length of a curve as $\int \dd \tau \sqrt{2 g_2} $, where $g_2 \equiv \frac12 g_{\mu \nu} \dot x^\mu \dot x^\nu$, one considers a more general $\int \dd \tau F(x, \dot x)$. The homogeneity rule $F(x, \lambda \dot x)= \lambda F(x,\dot x)$ is needed to ensure that such a length does not depend on how the curve is parameterized, but this still leaves many possibilities. A homogeneous analogue of a power series leads to a fairly general expression:\footnote{Odd $s$ can also be introduced, as we will discuss later.}
\begin{equation}\label{eq:F-xdot}
	F^2 = 2g_2 + \frac{\phi_4}{g_2} + \frac{\phi_6}{g_2^2}+ \ldots \, ,\qquad \phi_s \equiv \frac1{s!}\phi_{\mu_1 \ldots \mu_s} \dot x^{\mu_1} \ldots \dot x^{\mu_s}\,.
\end{equation}
So the first term in this expansion leads to the usual notion of distance; the others are new. 
 One can also think of this as a line element $\dd s^2= \dd s^2_0 $ $+ (2/\dd s^2_0) \phi_{\mu \nu \rho \sigma} \dd x^\mu \dd x^\nu \dd x^\rho \dd x^\sigma $ $+\ldots$, where $\dd s^2_0= g_{\mu \nu}\dd x^\mu \dd x^\nu$ is of the customary (pseudo-)Riemannian type. 

Analogues of the usual notions of connection and curvature have been studied for Finsler geometry for a long time. These notions are most natural when $F$ is considered as a function on the tangent bundle $TM$,\footnote{More precisely, because of the homogeneity constraint on $F$, one has to work either with the \emph{slit} tangent bundle $TM-$(zero section), or with the \emph{sphere} bundle $SM$ inside it, where each fiber is quotiented by overall rescalings.} with $\dot x^\mu$ now promoted to a coordinate $y^\mu$ along the fiber.

Mathematically it looks natural to build a Finsler modification of general relativity (GR); this has indeed been explored at length in the literature (see for example \cite{lammerzahl-perlick,pfeifer-review} for reviews). But from a physical point of view, no model of gravity can be considered to be well-motivated unless it improves on Einstein gravity on the crucial issue of its quantum behavior. This important issue appears to have been relatively unexplored. 

In any case, the appearance of the symmetric tensors $\phi_{\mu_1 \ldots \mu_s}$ in (\ref{eq:F-xdot}) suggests a relation to higher-spin theories. Given the promising status of the latter as quantum theories, such a link would also make Finsler geometry a lot more interesting physically.  
The $\int \dd \tau F$ would then be interpreted as a natural coupling of the $\phi_{\mu_1 \ldots \mu_s}$ to a particle. 

A version of (\ref{eq:F-xdot}) with $s=2$ and one $s>2$ was considered in this context \cite[(5.8)]{dewit-freedman}, and an approximate gauge transformation for the particle action was noticed. This was further explored and generalized in \cite{segal,ponomarev-symplectic}, in a line of research that led to a conformal theory of higher spins \cite{segal-chs}.\footnote{These references use a Hamiltonian point of view, which was recently further explored in \cite{ivanovskiy-ponomarev,ivanovskiy-ponomarev2} in the context of self-dual higher spins \cite{ponomarev-skvortsov}. I thank D.~Ponomarev for insightful comments.}
In \cite{hull-W} an analogue of Finsler geometry (without the homogeneity constraint) was used for W-gravity in low dimensions. Most relevant for us, (\ref{eq:F-xdot}) was found in \cite{guo} for $s=4$ and $g_2=\eta_2$ (flat space) to be related to the spin-4 Fronsdal equation \cite{fronsdal}.

In this paper I will explore (\ref{eq:F-xdot}) more systematically, in the spirit of taking the $\phi_s$ to be Finsler deformations around an ordinary (pseudo-)Riemannian geometry described by $g_2=\frac12 g_{\mu \nu}y^\mu y^\nu$. 
The first intriguing result is that the Fronsdal kinetic operator ${\mathcal F}(\phi_s)$ appears for all spins $s>2$ and around any $g_2$. Among the curvature tensors, one that we will call $\rho$ is the analogue of the ordinary Ricci tensor. We will see that
\begin{equation}\label{eq:Fgs-d2-intro}
	\rho \sim R^0_{\alpha \beta} y^\alpha y^\beta
	 + \sum_{s>2} g_2^{1-s/2} \left(-\frac12{\mathcal F}(\phi_s) + \alpha_s \dd^2 \phi_{s-2}\right)\,,
\end{equation}
to linear order in the $\phi_s$.
The first term contains the ordinary Ricci tensor of $g_2$; the rest is a series similar to (\ref{eq:F-xdot}). In a condensed notation to be fully explain below, $\dd$ represents the symmetrized derivative $\nabla_{(\mu_1} \phi^{s-2}_{\mu_2 \ldots \mu_s)}$. The coefficient $\alpha_s =\frac18 (s-4)(s+D-4)$, where $D$ is the spacetime dimension, which we will leave unspecified throughout. A perhaps deeper expression is 
\begin{equation}\label{eq:R-def-intro}
	\rho\sim R^0_{\alpha \beta} y^\alpha y^\beta - \frac12 {\mathcal F}(\delta F^2)
\end{equation}
with an appropriate understanding of the action of ${\mathcal F}$ on arbitrary functions of $y$. This is valid for Finsler deformations $\delta F^2$ around any (pseudo-)Riemannian geometry, without even using the expansion (\ref{eq:F-xdot}). 

A second interesting point is that the Finsler Ricci tensor is not linear in the $\phi_s$; going to higher orders gives rise to expressions that are more complicated but still manageable, especially for low $s$. It is natural to interpret these as higher spin interactions, similar to the graviton interactions appearing in the perturbative expansion of GR around Minkowski space.

Unfortunately there is also bad news. What makes the Fronsdal operator especially important is that it admits gauge transformations: under $\delta \phi_s = \dd \lambda_{s-1}$, $\delta{\mathcal F}(\phi_s)= \frac12 \dd^3 \lambda'_{s-1}$, with $\lambda'_{s-1}$ representing the trace $\lambda^\mu{}_{\mu \mu_2\ldots \mu_s}$. Thus the traceless $\lambda_{s-1}$ give a large set of gauge transformations.\footnote{This traceless condition is often viewed with suspicion; there is a way to get rid of it \cite{francia-sagnotti}. In the context of Finsler geometry we will see that in a sense this condition is rather natural.} On the other hand, the double symmetrized derivative $\dd^2$ in (\ref{eq:Fgs-d2-intro}) is not invariant. The two terms combine to transform as $\delta({\mathcal F}(\phi_s) - 2 \alpha_s \dd^2 \phi_{s-2})= \dd^3(-\frac14 \lambda'_{s-1}+ \alpha_s \lambda_{s-3})$. The gauge transformation of each field is ``eaten'' by that of a higher spin, in a sort of St\"uckelberg mechanism. Unfortunately, the surviving transformations can be shown to be a trivial rearrangement of fields that leaves the geometry unchanged. (An equivalent, sleeker analysis is possible from (\ref{eq:R-def-intro}).) 

This essential lack of gauge transformations (beyond the usual diffeomorphisms) is worrisome. We will see that the equations of motion for a Finsler analogue of GR, the simplest of which just reads $\rho=0$, have many perturbative solutions that are not transverse to the momentum, and thus would presumably create trouble upon quantization. These solutions also have many free parameters; this might give hope that a gauge transformation can indeed be defined for this system.  Unfortunately I have found no such candidate, but I have not excluded it either. These negative results are not entirely unexpected, in light of some of the no-go results mentioned above. The spirit of this paper is that reexamining an issue in a new light can lead to new insights. Overall, the idea of a Finsler higher spin theory faces significant challenges. Nevertheless, in my opinion the present results make it worth exploring further in the future.

In Sec.~\ref{sec:fronsdal} some aspects of higher spin field theory are reviewed, setting up notation. Sec.~\ref{sec:finsler} is a lightning introduction to Finsler geometry. In Sec.~\ref{sec:ff} I connect the two, showing that a linearization of the Finsler analogue $\rho$ of the Ricci tensor contains infinitely many Fronsdal operators, as promised above. In Sec.~\ref{sec:nonlin} a partial analysis is given beyond the linear order in the $\phi_s$. Sec.~\ref{sec:challenges} considers the physical issues of gauge transformations and of perturbative solutions, with some details in App.~\ref{app:d}.


\section{Fronsdal equation} 
\label{sec:fronsdal}

The Fronsdal equation \cite{fronsdal} describes a massless free field in a totally symmetric representation. We will give here a very basic review of some its features. There are many deeper discussions in the literature \cite{sagnotti-notes-hs,sorokin-hs,ponomarev-hs}.

\subsection{Spin two} 
\label{sub:spin2}

As a warm-up, let us recall the equation for a spin-two field. This can be obtained by applying the vacuum equations of general relativity, $R_{\mu \nu}=0$, to an infinitesimal perturbation $g_{\mu \nu}+ h_{\mu \nu}$ of a background metric $g_{\mu \nu}$. The connection is deformed by
\begin{equation}\label{eq:G0-def}
	\delta \Gamma^\mu_{\nu \rho}= g^{\mu \sigma} \left(\nabla_{(\nu} h_{\rho) \sigma}-\frac12 \nabla_\sigma h_{\nu \rho}\right)
\end{equation}
and the Riemann tensor by
\begin{equation}\label{eq:Ricci0-def}
	\delta R_{\nu \rho}= 2\nabla_{[\mu}\delta \Gamma^{\mu}_{\nu] \rho}=
	-\frac12 \nabla^2 h_{\nu \rho}+\nabla^\mu \nabla_{(\nu} h_{\rho) \mu} -\frac12 \nabla_\nu \nabla_\rho h'\,,
\end{equation}
where $h'= g^{\mu \nu} h_{\mu \nu}$. So setting the right-hand side to zero is the equation of motion for $h_{\mu \nu}$.

Sometimes it can be useful to rewrite (\ref{eq:Ricci0-def}) in terms of the \emph{Lichnerowicz operator} $\Delta$:
\begin{subequations}
\begin{align}
	&\delta R_{\nu \rho}= -\frac12 \Delta \delta g_{\nu \rho} + \nabla_{(\nu} \nabla^\mu \delta g_{\rho)\mu}-\frac12 \nabla_\rho \nabla_\nu \delta g' \, ,\\
	\label{eq:lichnerowicz}
	&\Delta \delta g_{\nu \rho}\equiv \nabla^2 \delta g_{\nu \rho}-2 R_{(\nu}{}^\sigma \delta g_{\rho) \sigma} +2 R_{\mu \nu \rho \sigma}h^{\mu \sigma}\,.	
\end{align}	
\end{subequations}

The gauge transformations of general relativity are infinitesimal coordinate changes; if we take
\begin{equation}\label{eq:deltaxi-h}
	\delta h_{\mu \nu}= 2\nabla_{(\mu} \xi_{\nu)}\,,
\end{equation}
(\ref{eq:Ricci0-def}) gives
\begin{equation}\label{eq:deltaxi-R}
	\delta_\xi R_{\mu \nu}= \xi^\rho \nabla_\rho R_{\mu \nu} +2 \nabla_{(\mu} \xi^\rho R_{\nu) \rho}= L_\xi R_{\mu \nu}
\end{equation}
after using the familiar
\begin{equation}\label{eq:[nn]}
	[\nabla_\mu, \nabla_\nu] v^\rho = R^\rho{}_{\sigma\mu \nu}v^\sigma
\end{equation}
and its generalization on tensors. If the original metric is a solution of the vacuum equation, $R_{\mu \nu}=0$, and so (\ref{eq:Ricci0-def}) is invariant.


\subsection{Symmetric products} 
\label{sub:sym}

To set the stage for a generalization to higher spins, let us consider now a condensed notation. A version of this formalism is used one way or another in much of the higher-spin literature; see for example \cite[(2.10)]{sagnotti-notes-hs}. My definitions are adapted to Finsler geometry, so some normalizations might be a bit unfamiliar. 

We introduce a formal variable $y^\mu$ (which will later acquire an interpretation as a velocity vector). To a completely symmetric tensor $\phi_{\mu_1\ldots \mu_s}$ we associate a polynomial
\begin{equation}\label{eq:phi-s}
	\phi_s \equiv \frac1{s!} \phi_{\mu_1 \ldots \mu_s} y^{\mu_1}\ldots y^{\mu_s}\,; 
\end{equation} 
this is of course similar to the notation in (\ref{eq:F-xdot}). It is convenient to also define
\begin{equation}\label{eq:phis-m}
\begin{split}
		\phi^s_\mu &\equiv \partial_{y^\mu} \phi_s = \frac1{(s-1)!} \phi_{\mu_2\ldots \mu_s} y^{\mu_2}\ldots y^{\mu_s}\,,\\
		\phi^s_{\mu \nu} &\equiv \partial_{y^\mu}\partial_{y^\nu} \phi_s = \frac1{(s-2)!} \phi_{\mu_3\ldots \mu_s} y^{\mu_3}\ldots y^{\mu_s}\,,
\end{split}
\end{equation}
and so on. The $s$ is a label and can be written up or down as convenient, and will sometimes be omitted when clear from the context. Notice that
\begin{equation}\label{eq:euler-y}
	y^\mu \phi^s_\mu = s\phi_s\,.
\end{equation}
The trace is defined as
\begin{equation}\label{eq:trace}
	\phi'_s \equiv \phi^\mu_{s\, \mu} = g^{\mu \nu}\partial_{y^\mu}\partial_{y^\nu} \phi_s\,.
\end{equation}

A product of two polynomials $\phi_s \psi_{s'}$ represents in this language the symmetric product $\frac{(s+s')!}{s! s'!}\phi_{(\mu_1\ldots \mu_s} \psi_{\mu_{s+1} \ldots \mu_{s+s'})}$. We will often encounter products $g_2 \phi_s$ or more generally $g_2^k \phi_s$, with $g_2= \frac12 g_{\mu \nu}y^\mu y^\nu$ and $g_{\mu \nu}$ a metric. It is useful to compute the trace of such objects: 
\begin{equation}\label{eq:(g2phi)'}
	(g_2 \phi_s)'=g^{\mu \nu}\partial_{y^\mu} (y_\nu \phi_s + g_2 \phi^s_\nu)= D \phi_s + 2 y^\mu \phi_\mu^s + g_2 \phi_s' = (D+2s) \phi_s + g_2 \phi'_s\,.
\end{equation}
One can rewrite this more abstractly by introducing the operators
\begin{equation}\label{eq:t-deg}
	t\equiv g^{\mu \nu} \partial_{y^\mu} \partial_{y^\nu}= \partial_{y^\mu} \partial_{y_\mu}
	\, ,\qquad
	\mathrm{deg} \equiv y^\mu \partial_{y^\mu}\,,
\end{equation}
so that $t \phi_s = \phi_s'$ and $\mathrm{deg} \phi_s = s \phi_s$, respectively. With the usual commutator rules, it is easy to work out
\begin{equation}\label{eq:[t,g2]}
	[t,g_2\cdot] = \partial_{y^\mu} y^\mu +y^\mu \partial_{y^\mu} = D+2\deg\,, 
\end{equation}
where $g_2\cdot$ is multiplication by $g_2$. This reproduces (\ref{eq:(g2phi)'}).

Any $\phi_s$ can be written in terms of traceless tensors:
\begin{equation}\label{eq:phi-dec}
	\phi_s= \phi_{s,0} + g_2 \phi_{s,1} + \ldots g_2^{s/2} \phi_{s,s} \, ,\qquad \phi_{s,k}'=0\,.
\end{equation}
The degree $(s-2k)$ fields $\phi_{s,k}$ can also be written in terms of the iterated traces $\phi^{(j)}$ (in a notation where $\phi^{(1)}= \phi'$, $\phi^{(2)}=\phi''$ and so on):
\begin{equation}\label{eq:phi-s0}
	\phi_{s,0}= \phi_s + t_{s1} g_2 \phi_s' + t_{s2} \frac{g_2^2}2 \phi_s''+ \ldots \, ,\qquad \phi_{s,j}=(-)^j t_{s-j+1,j} (\phi^{(j)})_0\,,
\end{equation}
where 
\begin{equation}\label{eq:tsj}
	t_{s1}\equiv-\frac1{D+2s-4} \, ,\qquad t_{s,j}\equiv - \frac{t_{s,j-1}}{D+2s-2j-2}\,.
\end{equation}


\subsection{Differential operators} 
\label{sub:diff}

The placeholder variable $y$ is inert under derivatives. So $\nabla_\mu \phi_s = \frac1{s!} \nabla_\mu\phi_{\mu_1 \ldots \mu_s} y^{\mu_1}\ldots y^{\mu_s}$. We will often encounter the operator
\begin{equation}\label{eq:d}
	\dd \equiv y^\mu \nabla_\mu\,.
\end{equation}
This name is inspired by $\dd = \dd x^\mu \wedge \partial_\mu= \dd x^\mu \wedge \nabla_\mu$ in exterior algebra, which represents an antisymmetrized derivative, similar to how (\ref{eq:d}) represents a \emph{symmetrized} derivative:
\begin{equation}
	\dd \phi_s = \frac1{s!}\nabla_{(\mu} \phi_{\mu_1 \ldots \mu_s)}y^\mu y^{\mu_1}\ldots y^{\mu_s}\,.
\end{equation}
(Writing the symmetrizer on the indices is here of course optional, as the product of $y$'s enforces it anyway.) A crucial difference is that the $\dd$ in exterior algebra squares to zero, while (\ref{eq:d}) does not.

In this language, the equation of motion $R_{\mu \nu}=0$ from (\ref{eq:Ricci0-def}), namely $\frac12 \nabla^2 h_{\nu \rho}-\nabla^\mu \nabla_{(\nu} h_{\rho) \mu} +\frac12 \nabla_\nu \nabla_\rho h'=0$, can be rewritten by multiplying by $y^\nu y^\rho$ as
\begin{equation}\label{eq:h-eom}
	\nabla^2 h_2 - \nabla_\mu \dd h_2^\mu +\frac12 \dd^2 h'=0\,.
\end{equation}
The gauge transformation (\ref{eq:deltaxi-h}) reads
\begin{equation}\label{eq:delta-h2}
	\delta h_2 = \dd \xi_1\,.
\end{equation}

We seem not to have gained much by this rewriting. However, the nice properties of (\ref{eq:h-eom}) now suggest to replace $h_2$ in (\ref{eq:h-eom}) by a polynomial $\phi$ in $y$, obtaining the equation of motion
\begin{equation}\label{eq:fronsdal}
	{\mathcal F}(\phi)\equiv \nabla^2 \phi - \nabla_\mu \dd \phi^\mu +\frac12 \dd^2 \phi'=0\,
\end{equation}
irrespectively of the degree of $\phi$. This is the \emph{Fronsdal equation} for massless totally symmetric fields. It is much simpler than its explicit expression: for $\phi=\phi_s$ as in (\ref{eq:phi-s}),
\begin{equation}\label{eq:fronsdal-exp}
	\nabla^2 \phi_{\mu_1 \ldots \mu_s} - s \nabla_\mu \nabla_{(\mu_1} \phi_{\mu_2 \ldots \mu_s)}{}^\mu +\frac{s(s-1)}2 \nabla_{(\mu_1} \nabla_{\mu_2} \phi_{\mu_3 \ldots \mu_s) \mu}{}^\mu=0 \,.
\end{equation}
We will see soon that this equation is invariant under a gauge transformation, but only in maximally symmetric spaces.

We can also rewrite 
\begin{equation}
	{\mathcal F}= \nabla^2- \nabla^\mu \dd \partial_{y^\mu} +\frac12\dd^2 t = \Delta - \dd \dd^\dagger +\frac12 \dd^2 t \,,
\end{equation}
with $\Delta= \nabla^2 - R_{\mu \nu}y^\mu \partial_{y_\nu} + R_{\mu \rho \nu \sigma} y^\mu y^\nu \partial_{y_\rho} \partial_{y_\sigma}$ as in (\ref{eq:lichnerowicz}), recalling (\ref{eq:t-deg}), and with the definition
\begin{equation}
	\dd^\dagger \equiv g^{\mu \nu} \partial_{y^\mu} \nabla_\nu= \partial_{y_\mu} \nabla_\mu \, .
\end{equation}
We note for later use:
\begin{equation}\label{eq:d-comm}
	[t,\dd]=2\dd^\dagger \, ,\qquad [t,\dd^\dagger] \, ,\qquad
	[\dd^\dagger,g_2\cdot]=\dd \, ,\qquad [\dd,g_2\cdot]=0\,,
	[\dd,\dd^\dagger]=-2 \nabla^2 + \Delta\,.
\end{equation}


\subsection{Flat space} 
\label{sub:flat}

We will now see that in flat space (\ref{eq:fronsdal}) inherits some of the nice properties of (\ref{eq:h-eom}). So in this subsection $g_{\mu \nu}= \eta_{\mu \nu}$. 

In flat space, a $\phi_s$ give a representation of the Lorentz group; it is irreducible if $\phi'_s=0$. Otherwise, (\ref{eq:phi-dec}) can be viewed as a decomposition into $s$ irreducible representations. 

Generalizing the gauge transformation (\ref{eq:delta-h2}), we take that of a field $\phi=\phi_s$ to be
\begin{equation}\label{eq:delta-phis}
	\delta \phi_s = \dd \lambda_{s-1}\,.
\end{equation}
In components this reads $\delta \phi_{\mu_1 \ldots \mu_s} = s \partial_{(\mu_1} \lambda_{\mu_2 \ldots \mu_s)}$. We find
\begin{equation}\label{eq:delta-gmu}
	\delta \phi^\mu = \partial_{y_\mu} (y^\alpha \partial_\alpha \lambda)= \partial^\mu \lambda + \dd \lambda^\mu
	\, ,\qquad
	\delta \phi^{\mu \nu} = 2 \partial^{(\mu} \lambda^{\nu)} + y^\alpha \partial_\alpha \lambda^{\mu \nu}\,.
\end{equation}
(As in (\ref{eq:fronsdal}), when the equations don't depend on the degree/spin we sometimes omit it, to get more readable expressions.)  
So for the Fronsdal equation (\ref{eq:fronsdal}):
\begin{equation}\label{eq:delta-Fs}
	\begin{split}
	\delta {\mathcal F}(\phi)&= 		{\mathcal F}(\dd \lambda)
		 =
		 \partial^2 \dd \lambda - \dd \partial_\mu (\partial^\mu \lambda + \dd \lambda^\mu) +\frac12 \dd^2(2 \partial_\mu \lambda^\mu + \dd \lambda')\\
	 &=\frac12 \dd^3 \lambda'\,.
	\end{split}
\end{equation}
In particular, ${\mathcal F}(\phi_s)$ is invariant under $\lambda_{s-1}$ that are traceless: 
\begin{equation}\label{eq:lambda-traceless}
	\lambda'= \lambda_{\alpha}^\alpha=0\,.
\end{equation}
Recall that more explicitly this means $\lambda^\alpha{}_{\alpha \mu_3\ldots \mu_{s-1}}=0$.

Trying to write an action for the Fronsdal equation leads to a second constraint. Up to integration by parts,  
\begin{equation}
	S = \int \dd^D x \,\phi\, {\mathcal G} \, ,\qquad  {\mathcal G} \equiv {\mathcal F}(\phi)-\frac12 \eta_2 {\mathcal F}'(\phi)\,.
\end{equation}
This action is in general not gauge invariant: $\delta S =\int \dd^D x \,\lambda\,\partial_\mu {\mathcal G}^\mu\neq 0$. Fortunately, a few more cancellations yield
\begin{equation}\label{eq:bianchi-G}
	\partial_\mu {\mathcal G}^\mu = -\frac14 \dd^3 \phi''\,.
\end{equation}
We see that the action is invariant if we also impose that the double trace of the field vanishes:
\begin{equation}\label{eq:phi''}
	\phi''=\phi^{\alpha \beta}{}_{\alpha \beta}= 0\,.
\end{equation}
(Recall that more explicitly this means $\phi^{\alpha \beta}{}_{\alpha \beta \mu_5\ldots \mu_s}=0$.) In this case $\phi$ consists of two irreducible representations, its trace $\phi'$ and its traceless part.

With this constraint, the variation $\delta S = \int \dd^D x \delta \phi {\mathcal G}=0$ now only imposes that the double-traceless part of ${\mathcal G}$ vanishes. Fortunately, it is simple to see (just by counting the number of available indices) that (\ref{eq:phi''}) in fact directly implies ${\mathcal F}''={\mathcal G}''=0$. So the action does imply ${\mathcal G}=0$. This in turn sets to zero ${\mathcal G}'$, which is proportional to ${\mathcal F}'$. So the equation of motion is indeed ${\mathcal F}=0$.

For spin two, the first constraint  (\ref{eq:lambda-traceless}) was automatic because in that case $\lambda_1$ was in fact $\xi_\mu$, which only has one index. (\ref{eq:bianchi-G}) becomes the linearized flat-space version of the spin-two identity $\nabla^\mu (R_{\mu \nu}-\frac12 R g_{\mu \nu})=0$, a consequence of the Bianchi identity. The double-trace constraint (\ref{eq:phi''}) is also automatic, since the field only has two indices.

We conclude with a quick review of gauge fixing. First we notice
\begin{equation}
	\delta \left(\partial_\mu \phi^\mu -\frac12 \dd \phi'\right)= \partial^2 \lambda \, ,\qquad 
	\left(\partial_\mu \phi^\mu -\frac12 \dd \phi'\right)'= -\frac12 \dd \phi''=0\,.
\end{equation}
So we can always find a $\lambda$ such that $\partial_\mu \phi^\mu -\frac12 \dd \phi'=0$, by inverting the operator $\partial^2$. This simplifies the Fronsdal equation to $\partial^2 \phi=0$. There is still a residual gauge invariance, consisting of $\lambda$ such that $\partial^2 \lambda=0$ (and which are traceless). 

We now go to momentum space, where we have $p^2 \phi(p)=0$, $p_\mu \phi^\mu = \frac12 p_1 \phi'$, with $p_1=y^\mu p_\mu$ as by now familiar. We pick a vector $u^\mu$ such that $u^\mu p_\mu=1$. Using (\ref{eq:delta-gmu}),
\begin{equation}
	\delta(u^\mu \phi_\mu ) = \lambda+ u_\mu y^\alpha p_\alpha \lambda^\mu = M \lambda \, ,\qquad M\equiv 1 + p_1 u^\mu \partial_{y^\mu}\,.
\end{equation}
It can be shown that $M$ is invertible, with inverse $M^{-1}= \sum_l (-)^l (p_1^l/l!) \partial_{y^{\mu_1}}\ldots \partial_{y^{\mu_l}}$. Moreover $M^{-1}(-u^\mu \phi_\mu)= \sum_l (-)^l (p_1^l/l!) u_{\mu_1}\ldots u_{\mu_l} \phi^{\mu_1\ldots \mu_l}$ can be shown to be traceless. So by picking this $\lambda$ we can set $u^\mu \phi_\mu$ to zero. Since $M$ is invertible, it has no kernel, and so there is no residual gauge transformation. Finally, $p_\mu \phi^\mu = \frac12 p_1 \phi'$ implies $2p^\nu \phi_{\mu \nu} = p_\mu \phi'+p_1 \phi'_\mu$; contracting this with $u_\mu$ we find now that $\phi'=0$, and so also that $p_\mu \phi^\mu=0$. This is the analogue of the transverse traceless gauge for spin-two fields.


\subsection{Curved space} 
\label{sub:curved}

In curved space, the gauge transformation of $\phi_s$ is still (\ref{eq:delta-phis}), but it now represents $\delta \phi_{\mu_1 \ldots \mu_s} = s \nabla_{(\mu_1} \lambda_{\mu_2 \ldots \mu_s)}$, with covariant derivatives. The transformation of (\ref{eq:fronsdal}) is more complicated: some terms that canceled in (\ref{eq:delta-Fs}) no longer do so, because of the non-commutativity of covariant derivatives recalled in (\ref{eq:[nn]}).

Besides the explicit indices, one also needs to take care of the tensor indices inside $\phi_s$, made explicit in (\ref{eq:phi-s}). In this respect, the following identity is useful:
\begin{equation}
	[\nabla_\mu, \nabla_\nu] \phi= - R^\rho{}_{\sigma\mu \nu} y^\sigma \partial_{y^\rho} \phi\,.
\end{equation}

Using also the second Bianchi identity we obtain
\begin{equation}\label{eq:delta-Fs-curved}
	\delta {\mathcal F}(\phi)=\frac12 \dd^3 \lambda' -y^\alpha y^\beta (\lambda^\nu \nabla_\nu R_{\alpha \beta} + 2 \nabla_{(\alpha} \lambda^\nu R_{\beta) \nu})
		+ \dd \rho_{\mu \nu} \lambda^{\mu \nu} +2 \rho_{\mu \nu}\dd \lambda^{\mu \nu}\,.  
\end{equation}
We have introduced 
\begin{equation}\label{eq:rho-riem}
	\rho_{\mu \nu} \equiv R_{\mu \alpha \nu \beta} y^\alpha y^\beta\,
\end{equation}
which will be useful later on. The parenthesis with the Ricci terms in (\ref{eq:delta-Fs-curved}) is reminiscent of the Lie derivative $L_\xi R_{\alpha \beta}= \xi^\mu \nabla_\mu R_{\alpha \beta}+ 2\nabla_{(\alpha} \xi^\nu R_{\beta) \nu}$, except that $\lambda^\mu= \lambda^\mu_{s-1}= \frac1{(s-1)!}\lambda^\mu{}_{\mu_2 \ldots \mu_{s-1}}$ has $(s-2)$ hidden indices. For $s=2$ this expression does reduce to the Lie derivative in (\ref{eq:deltaxi-R}).

The appearance of the Riemann tensor in (\ref{eq:delta-Fs-curved}) is famously problematic \cite{aragone-deser}. We can still take $\lambda'=0$ to make the first term disappear. The Ricci tensor appears in the ordinary gravity equations, and in particular it simply vanishes for vacuum GR, so those terms can also be reasonably set to zero. However, setting the Riemann tensor to any particular value appears unjustified. So the gauge invariance of the Fronsdal equation is broken in general. 

If one simply postulates that the background geometry is maximally symmetric, as in AdS$_D$ or dS$_D$, then
\begin{equation}
	R_{\mu \nu \rho \sigma}= \frac{\Lambda}{D-1}(g_{\mu \rho} g_{\nu \sigma}- g_{\mu \sigma} g_{\nu \rho})\,.
\end{equation}
The constant is chosen so that $R_{\mu \nu}= \Lambda g_{\mu \nu}$. (\ref{eq:delta-Fs-curved}) now becomes the simpler
\begin{equation}\label{eq:deltaF-AdS}
	\delta {\mathcal F}(\phi_s)=\frac12 \dd^3 \lambda'_{s-1} + \frac{4 \Lambda}{D-1} g_2 \dd \lambda'_{s-1} - \Lambda\mu_s^2 \dd \lambda_{s-1} 
	\, ,\qquad \mu_s^2 \equiv \frac{2(s-1)(D+s-3)}{D-1}\,.
\end{equation}
(We reinstated the spin labels because now the equation does explicitly depend on $s$.) Recall that $g_2=\frac12 g_{\mu \nu} y^\mu y^\nu$, per our notation (\ref{eq:phi-s}). This is still non-zero even if $\lambda'=0$. But the modified equation 
\begin{equation}
	{\mathcal F}(\phi_s) + \Lambda\mu_s^2 \phi_s=0
\end{equation}
does have a gauge invariance.



\section{Finsler geometry} 
\label{sec:finsler}

We will give here a relatively quick introduction to Finsler geometry. Some relatively recent accounts are \cite{chern-shen,bao-chern-shen,shen-shen}, in increasing order of detail.

\subsection{Distance} 
\label{sub:distance}

In Riemannian geometry, the length $\gamma$ of a curve is given by the integral $\int_\gamma \dd \tau\sqrt{g_{\mu \nu}(x)\dot x^\mu \dot x^\nu}$, with $\tau$ a coordinate on $\gamma$ and $\dot x^\mu \equiv \partial_\tau x^\mu$. The idea of Finsler geometry is to generalize this to
\begin{equation}\label{eq:F-dist}
	\int \dd \tau F(x,\dot x)\,
\end{equation}
with $F(x,\dot x)$ any function of $x^\mu$, $\dot x^\mu$ that is homogeneous of degree one in velocity:
\begin{equation}\label{eq:F-hom}
	F(x, \lambda \dot x)= \lambda F(x, \dot x)\,.
\end{equation}
This requirement is needed so that the length (\ref{eq:F-dist}) of $\gamma$ is invariant under $\tau \to \tau'(\tau)$ and does not depend on how we parameterize it. Actually it is customary to call 
\begin{equation}
	\dot x^\mu \equiv y^\mu
\end{equation}
the velocity variable, and we will do so. This can be viewed as a coordinate along the fiber of the tangent bundle $TM$. In this sense, $F$ can be viewed as a function on $TM$, a fact that will be useful later. 

Clearly there are many new options opened by the generalization (\ref{eq:F-dist}), (\ref{eq:F-hom}). Two simple possibilities that have been considered in the mathematical literature are the \emph{Randers} choice $F= \phi_1 + \sqrt{2g_2}$ and the \emph{$s$-th root} choice $F=(\phi_s)^{1/s}$, with $\phi_s= \frac1{s!} \phi_{\mu_1\ldots \mu_s} y^{\mu_1}\ldots y^{\mu_s}$ as in (\ref{eq:phi-s}) and (\ref{eq:F-xdot}). That equation also shows infinitely many new possibilities, which will be explored in Sec.\ref{sec:ff} below. 

Most of the mathematical literature has been devoted to the case where $F$ is positive, thus defining a notion of distance that generalizes Riemannian geometry.\footnote{Not all metric spaces are of this form, as pointed out in \cite{tamassy}. There exist even more general notions of distance that cannot be obtained as integrals of a function $F(x,\dot x)$.} 
 In that case, $F$ is also taken to be smooth, and the \emph{fundamental tensor}
\begin{equation}\label{eq:g}
	g_{\mu \nu}  = \frac12 \partial_{y^\mu} \partial_{y^\nu} F^2
\end{equation}
is required to be positive-definite for all $x$ and $y$. When there is danger of confusion between (\ref{eq:g}) and a Riemannian metric, I will call the latter $g^0_{\mu \nu}$. Notice that (\ref{eq:g}) depends on $y$, except if $F^2=2g^0_2 = g^0_{\mu \nu}(x) y^\mu y^\nu$, where $g_{\mu \nu}= g^0_{\mu \nu}(x)$ is an ordinary Riemannian metric. (\ref{eq:g}) is used to lower indices in Finsler geometry; given that it is positive-definite, its inverse $g^{\mu \nu}$ is used to raise them.

For physics, we are more interested in generalizing pseudo-Riemannian spacetimes; for this, we require instead (\ref{eq:g}) to have signature $(-+++)$. Now in spite of its name $F^2$ is no longer assumed to be positive, similar to how the symbol $\dd s^2$ can have either sign in ordinary pseudo-Riemannian geometry. We take $\int \dd \tau\sqrt{F^2}$ to measure distances, and $\int \dd \tau \sqrt{-F^2}$ to measure proper time. 

Even with the understanding that $F^2$ can have either sign, some choices are not defined for all $y$. An example is the pseudo-Randers $F^2= \phi_1 + \sqrt{-2g_2}$, with $g_{\mu \nu}$ pseudo-Riemannian. Another example is encountered when generalizing (\ref{eq:F-xdot}) to include odd $s$, with terms $\phi_3/\sqrt{-g_2}+ \phi_5/(-g_2)^{3/2}+\ldots$. Physically this appears to be fine as long as, for every $x$, $F^2$ is real above a cone in the tangent space $T_x M$. Definitions in the mathematical literature vary as to whether to impose $F^2$ to be defined for all $y$ or not \cite{minguzzi-lightcones}. 
The rest of our discussion in this Section will be basically valid for either signature. 

Before we proceed further, we note that the homogeneity condition (\ref{eq:F-hom}) has several interesting consequences. Recall the \emph{Euler theorem} for a homogeneous function of degree $k$:
\begin{equation}
	\phi(\lambda y)= \lambda^k \phi(y) \quad \Rightarrow \quad y^\mu \partial_{y^\mu} \phi = k \phi\,.
\end{equation}
A monomial such as (\ref{eq:phi-s}) is obviously homogeneous of degree $s$, as we saw in (\ref{eq:euler-y}). If $\phi=\phi(x,y)$ is analytic around $y=0$, it can be Taylor expanded in such a basis: 
\begin{equation}\label{eq:taylor-y}
	\phi= \phi_0 + \phi_\mu y^\mu + \frac12 \phi_{\mu \nu} y^\mu y^\nu + \ldots\,.
\end{equation}
In this sense, such a function can be seen as a collection of completely symmetric tensors $\phi_{\mu_1\ldots \mu_s}$. 

$F^2$ is homogeneous of degree two; moreover, each $\partial_{y^\mu}$ lowers the homogeneity degree by one, so for example (\ref{eq:g}) has degree zero. This gives the following useful identities:
\begin{equation}\label{eq:yF}
	y^\mu \partial_{y^\mu} F^2 = 2 F^2 \, ,\qquad y^\mu g_{\mu \nu}= \frac12 \partial_{y^\nu} F^2 \, ,\qquad y^\mu y^\nu g_{\mu \nu} = F^2
	\, ,\qquad y^\mu \partial_{y^\mu} g_{\nu \rho}=0	\,.
\end{equation}


\subsection{Connection} 
\label{sub:conn}

Just like in ordinary (pseudo-)Riemannian geometry, a connection is needed in order to write derivatives that transform well under coordinate changes. A first piece of information is obtained by varying (\ref{eq:F-dist}): this gives the familiar-looking geodesic equation $\ddot x^\mu + \gamma^\mu_{\nu \rho}\dot x^\nu \dot x^\rho=0$ with 
\begin{equation}\label{eq:gamma}
	\gamma^\mu_{\nu \rho} = \frac12 g^{\mu \sigma} ( \partial_\nu g_{\rho \sigma}+ \partial_\rho g_{\nu \sigma} - \partial_\sigma g_{\nu \rho})\,.
\end{equation}
Recall however that $g$ now also depends on $y^\mu = \dot x^\mu$. The combination appearing in the geodesic equation appears often enough that we give it a name, \emph{spray coefficients}:
\begin{equation}\label{eq:G}
	G_\mu \equiv \gamma_{\mu \nu \rho} y^\nu y^\rho= 
	\frac12 (y^\nu \partial_\nu  \partial_{y^\mu} - \partial_\mu) F^2
	\,.
\end{equation}
The second equality uses (\ref{eq:yF}). Recall that we are using here the fundamental tensor (\ref{eq:g}) to lower indices.

We noticed earlier that $F$ is a function on $TM$. It proves useful to also think of the derivatives $\partial_\mu$ and $\partial_{y^\mu}$ as vector fields on that bundle. Under a coordinate change $x^\mu \to x'{}^\mu= x'{}^\mu(x)$, the velocity variables should transform as $y^\mu \to y'{}^\mu=\frac{\partial x'{}^\mu}{\partial x^\nu} y^\nu$. If we now view both formulas together as a coordinate change on $TM$, it follows that
\begin{subequations}
\begin{align}\label{eq:dex-transf}
	\partial_\mu &\to \partial'_\mu = \frac{\partial x'{}^\nu}{\partial x^\mu} \frac{\partial}{\partial x'{}^\nu} + \frac{\partial y'{}^\nu}{\partial x^\mu} \frac{\partial}{\partial y'{}^\nu} = \frac{\partial x'{}^\nu}{\partial x^\mu} \partial'_\nu + y^\rho\frac{\partial^2 x'{}^\nu}{\partial x^\mu \partial x^\rho} \partial_{y'{}^\nu}\,,
	\\
	\partial_{y^\mu} &\to \partial_{y'{}^\mu}= \frac{\partial y'{}^\nu}{\partial y^\mu} \frac{\partial}{\partial y'{}^\nu}=
	\frac{\partial x'{}^\nu}{\partial x^\mu} \partial_{y'{}^\nu} \,.	
\end{align}	
\end{subequations}
The $\partial_\mu$ transform in a complicated fashion, but it turns out that the combination 
\begin{equation}\label{eq:d-mu}
	\delta_\mu \equiv \partial_\mu - N^\nu{}_\mu \partial_{y^\nu} \, ,\qquad
	N^\nu{}_\mu \equiv \frac12 \partial_{y^\mu}G^\nu 
\end{equation}
transforms without the second term in (\ref{eq:dex-transf}). A dual issue appears when considering forms: on $TM$ the $\dd y^\mu$ transform in a complicated fashion, and this is solved by introducing 
\begin{equation}
	\delta y^\mu \equiv \dd y^\mu + N^\mu{}_\nu \dd x^\nu \,.
\end{equation}

To see the importance of (\ref{eq:d-mu}), consider the case of (pseudo-)Riemannian geometry, $F^2= 2g^0_2 = g^0_{\mu \nu} y^\mu y^\nu$. The fundamental tensor is $g_{\mu \nu}= g^0_{\mu \nu}$, and $N^\mu{}_\nu = \gamma^\mu_{\nu \rho}y^\rho$.\footnote{This is not true in general, because $\gamma^\mu_{\nu \rho}$ depends on $y$. By using (\ref{eq:yF}) one can show that $\partial_{y^\nu}G_\mu=2 \gamma_{\mu \nu \rho}y^\rho$, but $N^\nu{}_\mu$ contains the derivative of $G^\mu$ with an \emph{upper} index.} On each of the monomials $T_s= \frac1{s!} T_{\mu_1\ldots \mu_s} y^{\mu_1}\ldots y^{\mu_s}$ appearing in (\ref{eq:F-xdot}), we obtain $\delta_\mu T_s = \frac1{s!}\nabla_\mu T_{\mu_1\ldots \mu_s} y^{\mu_1}\ldots y^{\mu_s}$, where $\nabla$ is the covariant derivative relative to the (pseudo-)Riemannian metric $g^0_{\mu \nu}$. So for ordinary (pseudo-)Riemannian geometry $\delta_\mu $ coincides with the usual covariant derivative $\nabla_\mu$, even when acting on objects that depend on $y$.

Returning to the general discussion, the importance of (\ref{eq:d-mu}) suggests introducing the \emph{Chern connection}
\begin{equation}\label{eq:Gamma}
	\Gamma^\mu_{\nu \rho} = \frac12 g^{\mu \sigma} ( \delta_\nu g_{\rho \sigma}+ \delta_\rho g_{\nu \sigma} - \delta_\sigma g_{\nu \rho})\,.
\end{equation}
This has zero torsion, and satisfies
\begin{equation}\label{eq:Dg=0}
	D_\mu g_{\nu \rho} \equiv \delta_\mu g_{\nu \rho} -\Gamma^\sigma_{\mu \nu} g_{\sigma \rho}
	- \Gamma^\sigma_{\mu \rho} g_{\nu \sigma} = 0\,.
\end{equation}
We define the operator $D_\mu$ on any tensor, with $\Gamma$ acting on the indices in the way that is familiar in (pseudo-)Riemannian geometry, and the partial derivative replaced by $\delta$:\footnote{In the literature this derivative is usually denoted by a ``slash'' at the end: $T^{\mu_1\ldots \mu_k}{}_{\nu_1\ldots \nu_l| \mu}$.}
\begin{equation}\label{eq:covdevT}
\begin{split}
	D_\mu T^{\mu_1\ldots \mu_k}_{\nu_1\ldots \nu_l} = \delta_\mu T^{\mu_1\ldots \mu_k}_{\nu_1\ldots \nu_l}
	&+\Gamma^{\mu_1}_{\mu \rho}T^{\rho\mu_2\ldots \mu_k}_{\nu_1\ldots \nu_l}+ \Gamma^{\mu_2}_{\mu \rho}T^{\mu_1 \rho \mu_3 \ldots \mu_k}_{\nu_1\ldots \nu_l} + \ldots +\Gamma^{\mu_k}_{\mu \rho}T^{\mu_1\ldots \mu_{k-1} \rho}_{\nu_1\ldots \nu_l}\\
	&-\Gamma^\rho_{\mu \nu_1}T^{\mu_1\ldots \mu_k}_{\rho \nu_2\ldots \nu_l}- \Gamma^\rho_{\mu \nu_2}T^{\mu_1 \ldots \mu_k}_{\nu_1 \rho \nu_3\ldots \nu_l} - \ldots- \Gamma^\rho_{\mu \nu_l}T^{\mu_1\ldots \mu_k}_{\nu_1\ldots \nu_{l-1}\rho}\,.
\end{split}
\end{equation} 
Once again, this reduces to $\nabla_\mu$ in (pseudo-)Riemannian geometry, even if $T$ depends on $y$. In this sense, (\ref{eq:Dg=0}) is more natural than the analogous equality containing $\gamma^\mu_{\nu \rho}$ and ordinary partial derivatives.

At a more conceptual level, on $TM$ there is no connection that preserves the metric and has zero torsion; this is unlike in (pseudo-)Riemannian geometry, where (\ref{eq:gamma}) is the Levi-Civita connection and does have both properties.  (\ref{eq:gamma}) and (\ref{eq:Gamma}) both have zero torsion, but both fail to respect the metric in the $y$ directions of $TM$. If on $TM$ we write the exterior differential $\dd_{TM} \equiv \dd x^\mu \partial_\mu + \dd y^\mu \partial_{y^\mu}= \dd x^\mu \delta_\mu + \delta y^\mu \partial_{y^\mu}$, (\ref{eq:Dg=0}) implies 
\begin{equation}
	\dd_{TM} g_{\mu \nu}= \delta y^\rho \partial_{y^\rho} g_{\mu \nu}= \frac12\delta y^\rho \partial_{y^\mu} \partial_{y^\nu} \partial_{y^\rho} F^2\,.
\end{equation}
This property is called \emph{almost compatibility} with the metric of the Chern connection (\ref{eq:Gamma}). Other connections exist with nice properties, the most notable being the Cartan connection, which also has components along the $y$ coordinates; it is metric-compatible but has torsion \cite[Thm.~1.4.2]{abate-patrizio}.

The coefficients we introduced previously are related to the Chern connection:
\begin{equation}
	\Gamma^\mu_{\nu \rho}y^\rho = N^\mu{}_\nu \, ,\qquad
	N^\mu{}_\nu y^\nu = G^\mu\,.
\end{equation}
The second is a simple consequence of $G^\mu$ being homogeneous in $y$ with degree two; the first is a little more involved. 

It is also useful to notice that
\begin{equation}\label{eq:landsberg}
	[\partial_{y^\mu}, \delta_\nu] = - \partial_{y^\mu} N^\rho{}_\nu \partial_{y^\rho} = - (L^\rho_{\mu \nu} + \Gamma^\rho_{\mu \nu}) \partial_{y^\rho}\,,\qquad L^\rho_{\mu \nu} \equiv y^\sigma \partial_{y^\mu} \Gamma^\rho_{\nu \sigma}\,.
\end{equation}
We introduced the \emph{Landsberg tensor} $L^\rho_{\mu \nu}$. As an application,
\begin{equation}\label{eq:dy-delta}
	\partial_{y^\mu} \delta_\nu T = D_\nu \partial_{y^\mu} T - L^\rho_{\mu \nu} \partial_{y^\rho} T\,
\end{equation}
on a function $T=T(x,y)$. One can show that
\begin{equation}
	y^\nu L^\rho_{\mu \nu}= 0 \,,
\end{equation}
so in particular (\ref{eq:dy-delta}) implies
\begin{equation}\label{eq:d-delta}
	y^\nu \partial_{y^\mu} \delta_\nu T = y^\nu D_\nu \partial_{y^\mu} T = \dd \partial_{y^\mu} T\,.
\end{equation}
In the last step we have defined
\begin{equation}\label{eq:d-F}
	\dd \equiv y^\mu D_\mu\,
\end{equation}
by analogy with the $\dd$ in (\ref{eq:d}), to which it reduces in the (pseudo-)Riemannian case.
 
The contraction
\begin{equation}\label{eq:mean-landsberg}
	L^\mu = g^{\nu \rho} L^\mu_{\nu \rho} \, 
\end{equation}
is called \emph{mean} Landsberg tensor. On the other hand,
\begin{equation}
	L^\mu_{\mu \nu}=y^\rho \partial_{y^\nu}\Gamma^\mu_{\mu \rho}= \frac12 y^\rho \partial_{y^\nu} \delta_\rho \log \det g =\dd I_\nu\,,
\end{equation} 
where 
\begin{equation}\label{eq:mean-cartan}
	I_\mu = \frac12 \partial_{y^\mu} \log \det g\,, 
\end{equation}
is called the \emph{mean Cartan} tensor.


\subsection{Curvature} 
\label{sub:curv}

In view of (\ref{eq:Gamma}), we now define 
\begin{equation}\label{eq:riemann}
	\frac12R^\mu{}_{\nu \rho \sigma}\equiv \delta_{[\rho} \Gamma^\mu_{\sigma] \nu}+ \Gamma^\mu_{[\rho| \lambda} \Gamma^\lambda_{|\sigma] \nu}\,.
\end{equation}
It has many of the usual properties of the ordinary Riemann tensor: $R^{\mu}{}_{\nu (\rho \sigma)}=0$, $R^\mu{}_{[\nu \rho \sigma]}=0$. However, it is not antisymmetric in the first two indices: rather, it is possible to show
\begin{equation}\label{eq:riemann-notas}
	R_{(\mu \nu) \rho \sigma}= -\frac12 \partial_{y^\alpha} g_{\mu \nu} R^\alpha{}_{\beta \rho \sigma } y^\beta\,.
\end{equation}
A particular contraction of (\ref{eq:riemann}) with the $y$ can be reexpressed in terms of $N^\mu{}_\nu$ or of $G^\mu$:
\begin{subequations}\label{eq:R}
\begin{align}
	\label{eq:R-N}
	\rho^\mu{}_\rho &\equiv R^\mu{}_{\alpha \rho \beta} y^\alpha y^\beta = 2y^\sigma \delta_{[\rho} N^\mu{}_{\sigma]}\\
	\label{eq:R-G}
	  &= \partial_\rho G^\mu -\frac14 \partial_{y^\rho} G^\nu \partial_{y^\nu} G^\mu -\frac12 y^\nu \partial_\nu \partial_{y^\rho} G^\mu + \frac12 G^\nu \partial_{y^\nu} \partial_{y^\rho} G^\mu\,.
\end{align}
\end{subequations}
(\ref{eq:riemann-notas}) implies that $\rho_{\mu \rho}= \rho_{\rho \mu}$, $\rho_{\mu \rho}\equiv g_{\mu \nu}\rho^\nu{}_\rho$. We already encountered $\rho_{\mu \nu}$ in (\ref{eq:rho-riem}) in the (pseudo-)Riemannian case.\footnote{In the literature on Finsler geometry, $\rho^\mu{}_\nu$ is called $(1/F^2)R^\mu{}_\nu$. I have decided to change this notation in order to avoid confusion with the Ricci tensor in (pseudo-)Riemannian geometry. Even more confusingly, $\rho^\mu{}_\mu$ is often called ``Ricci scalar''; we will not follow this custom.}
The trace contains the Finsler extension of the Ricci tensor:
\begin{equation}\label{eq:ricci-finsler}
	\rho\equiv\rho^\mu{}_\mu = R^0_{\mu \nu} y^\mu y^\nu \, \qquad
	 (\text{when } F^2=g^0_{\mu \nu}y^\mu y^\nu)\,.
\end{equation}



\section{Linearized analysis} 
\label{sec:ff}

We would now like to obtain some formulas for infinitesimal deformations of Finsler structures, inspired by (\ref{eq:G0-def}), (\ref{eq:Ricci0-def}) in the ordinary (pseudo-)Riemannian case. As anticipated, this will lead to the appearance of the Fronsdal operator.

\subsection{Deformations} 
\label{sub:def}

We first take a small detour and show a formula for a \emph{finite} deformation. Given a $F^2$, consider a second Finsler structure $\tilde F^2$. It determines a new fundamental tensor, connection and curvature; we will denote them all with a tilde. After recalling (\ref{eq:d-mu}), one finds
\begin{equation}\label{eq:tG}
	\tilde G^\mu = G^\mu + \frac12 \tilde g^{\mu \rho} (y^\nu \partial_{y^\rho} \delta_\nu - \delta_\rho) \tilde F^2\,.
\end{equation}
It can be shown \cite[Thm.~3.3.1]{chern-shen} that the geodesics of $\tilde F^2$ coincide with those of $F^2$ iff $\delta_\nu \tilde F = y^\mu \partial_{y^\nu} \delta_\mu \tilde F$; in this case $\tilde G^\mu = G^\mu + P y^\mu $, with $P=\tilde F^{-1} y^\nu  \delta_\nu \tilde F$. The two Finsler $\tilde F$ and $F$ are said to be \emph{projectively equivalent}.

Coming now to the case of an infinitesimal deformation, (\ref{eq:tG}) gives
\begin{equation}\label{eq:G-def}
	\delta G^\mu = \frac12 g^{\mu \rho}(y^\nu \partial_{y^\rho} \delta_\nu - \delta_\rho) \delta F^2=
	\frac12 g^{\mu \rho}(\dd \partial_{y^\rho} - \delta_\rho) \delta F^2\,.
\end{equation}
In the second step we have used (\ref{eq:d-delta}), (\ref{eq:d-F}). ($\delta F^2$ is to be understood as $\delta(F^2)$.)

Recall from (\ref{eq:ricci-finsler}) that $\rho=\rho^\mu{}_\mu$ is the analogue of the Ricci tensor. So to obtain the analogue of (\ref{eq:Ricci0-def}) we need $\delta \rho= \delta \rho^\mu{}_\mu$. We can use either (\ref{eq:R-N}) or (\ref{eq:R-G}) to obtain
\begin{equation}\label{eq:R-def0}
	\delta  \rho = D_\mu \delta G^\mu +L^\mu_{\mu \nu}\delta G^\nu - \frac12 \dd \partial_{y^\mu} \delta G^\mu\,.
\end{equation}
Combining with (\ref{eq:G-def}):
\begin{equation}\label{eq:R-def}
\begin{split}
	\delta \rho=&\frac12g^{\mu \nu}\left( D_\mu \dd \partial_{y^\nu} - D_\mu \delta_\nu  - \frac12 \dd^2 \partial_{y^\mu} \partial_{y^\nu} \right) \delta F^2 \\
	&- \frac12\dd (L^\rho \partial_{y^\rho} \delta F^2) +\frac32 \dd I_\nu \delta G^\nu +\frac12 I_\mu \dd \delta G^\mu\,.
\end{split}
\end{equation}
(\ref{eq:R-def0}) and (\ref{eq:R-def}) are similar to the first and second step in (\ref{eq:Ricci0-def}). 

These results hold for a general deformation of a Finsler structure $F^2$. If the latter is in fact a (pseudo-)Riemannian geometry, (\ref{eq:G-def}) becomes
\begin{equation}\label{eq:G-def-riem}
	\delta G^\mu= \frac12 (\dd \partial_{y_\mu} - \nabla^\mu) \delta F^2\,.
\end{equation}
As for (\ref{eq:R-def}), since neither $g$ nor $\Gamma^\rho_{\mu \nu}$ depend on $y$, we have $L^\rho_{\mu \nu}=0=I_\mu$, so the second line in (\ref{eq:R-def}) vanishes. The last term on the first line contains $g^{\mu \nu}\partial_{y^\mu}\partial_{y^\nu}$; we saw back in (\ref{eq:trace}) that this gives the trace operator $t$ on monomials such as (\ref{eq:phi-s}), and now we can consider it as extending it on arbitrary (even non-polynomial) functions of $y$. This point will be crucial later. With this understanding, the first line of (\ref{eq:R-def}) is nothing but the Fronsdal operator (\ref{eq:fronsdal}):
\begin{equation}\label{eq:R-def-riem}
	\delta \rho = -\frac12 {\mathcal F} (\delta F^2)\,,
\end{equation}
as anticipated in the introduction. 

In the next subsection we will motivate the parameterization (\ref{eq:F-xdot}), and apply it to (\ref{eq:R-def-riem}). 


\subsection{Parametrization} 
\label{sub:param}

We will now justify the expansion (\ref{eq:F-xdot}). The idea is to start from a (pseudo-)Riemannian geometry, and to deform it into the more general Finsler type.

(\ref{eq:F-xdot}) is not quite a series expansion, since $F^2$ should be homogeneous. In Euclidean signature, one way to deal with this constraint is to think of it as a function on a sphere. For fixed $x$, $F^2(x,y)$ is a function on $S^{D-1}\cong\mathbb{R}^D/\mathbb{R}_+$, where the quotient acts by rescaling. A basis of functions on the round sphere is given by the spherical harmonics. These can be viewed as polynomials of the type $\phi_s$ in (\ref{eq:phi-s}) restricted to $\{2\delta_2\equiv\delta_{\mu \nu}y^\mu y^\nu=1\}$, or alternatively as functions on $\mathbb{R}^D$ made homogeneous of degree \emph{zero} by dividing them by an appropriate power of $\delta_2$:
\begin{equation}\label{eq:hom-phis}
	\frac1{s!}\frac1{\delta_2^{s/2}}\phi_{\mu_1 \ldots \mu_s}\,
\end{equation}
where $s$ is any non-negative integer, and the tensors $\phi_{\mu_1\ldots \mu_s}$ are irreducible: $\phi'_s=0$. 

This suggests that a good parameterization for $F^2$ can be obtained generalizing (\ref{eq:hom-phis}) by replacing $\delta_2$ with $g_2=\frac12 g^0_{\mu \nu}y^\mu y^\nu$. More precisely, $F^2$ is homogeneous of degree two, not zero; this can be repaired by multiplying by an overall $g_2$. This leads us to 
\begin{equation}\label{eq:F-y}
	F^2= 2 g_2 + \frac{\phi_3}{\sqrt{g_2}} + \frac{\phi_4}{g_2}+\ldots = 2 g_2 + \sum_{s>2} g_2^{1-s/2} \phi_s\,.
\end{equation} 
This extends (\ref{eq:F-xdot}) by including odd $s$. Actually their presence creates a possible issue in the pseudo-Lorentzian case because of the roots, as we discussed at the end of Sec.~\ref{sub:distance}. It becomes more sensible then to make $F^2$ defined above the light cone by changing $(g_2)^{s/2}\to (-g_2)^{s/2}$. Our discussion in the following is not influenced by this issue.

As mentioned earlier, the $\phi_s$ should be irreducible for an expansion that is complete and non-redundant. But no harm is done if the $\phi_s$ are not traceless and (\ref{eq:F-y}) is redundant. In that case, $F^2$ is trivially invariant (even at a finite level) under
	\begin{equation}\label{eq:tr-inv}
		\delta_\kappa \phi_{s_0}= \kappa_{s_0} \, ,\qquad \delta_\kappa \phi_{s_0+2}= - g_2 \kappa_{s_0}\,.
	\end{equation}
This redundancy will provide useful cross-checks of our results. It can be fixed to make the $\phi_s$ traceless again, or fixed only partially by enforcing a weaker constraint such as $\phi''_s=0$. (This was imposed in (\ref{eq:phi''}), but only in order to obtain a gauge-invariant action for (\ref{eq:fronsdal}); here we will not be using that action.) Alternatively one can proceed by expanding each $\phi_s$ in (\ref{eq:F-y}) in terms of its traceless components as in (\ref{eq:phi-dec}), collecting powers of $g_2$. This gives an expansion like (\ref{eq:F-y}) whose degree $s$ coefficient is now the traceless
\begin{equation}\label{eq:exp-proj}
	\sum_{j=0}^\infty \phi_{s+2j,j}\,.
\end{equation}
	
Similar to (\ref{eq:tr-inv}), there is an invariance where we deform
\begin{equation}
	\delta g_2 = \kappa_2 \, ,\qquad \delta \phi_s= (2-s/2) \kappa_2 g_{s-2}\,.
\end{equation}
One could even use this to connect higher spins around a curved space to higher spins around flat space, at the price of making the $\phi_s$ maximally non-irreducible.
	 
Strictly speaking, a complete parameterization would have included lower spins in (\ref{eq:F-y}); we have not done so because of our focus on higher spins. A spin-one term $\phi_1 \sqrt{g_2}$ term can be generated by changing $\phi_3 \to \phi_3 + g_2 \phi_1$ in (\ref{eq:F-y}), or in other words (\ref{eq:tr-inv}) for $s_0=1$. A spin-zero term $\phi_0 g_2$ is equivalent to rescaling the first term in (\ref{eq:F-y}), or (after renaming $g_2\to g_2(1+\phi_0)^{-1}$) to rescaling $\phi_s \to (1+\phi_0)^{s/2-1}$. In the future it might be interesting to investigate the effect of including this scalar; I have not done so in what follows. It would be especially intriguing to explore the limit in which $\phi_0\to-1$, and the first term in (\ref{eq:F-y}) would disappear.
	 
An alternative to (\ref{eq:F-y}) would be obtained by applying the same logic directly to $F$, leading to $F=\sqrt{2g_2} + f_3/g_2 + f_4/g_2^{3/2}+\ldots$. Adding the spin-one term to this parameterization leads to a generalization of the Randers choice mentioned in Sec.~\ref{sub:distance}.

We will use an expansion similar to (\ref{eq:F-y}) also for other $y$-dependent quantities, such as the connection and curvature:
\begin{equation}\label{eq:GR-y}
	G^\mu = \gamma^\mu_{0\, \alpha \beta}y^\alpha y^\beta + \sum_{s>2} g_2^{1-s/2} G^\mu_s
	\, ,\qquad
	\rho^\mu{}_\nu = R^\mu_{0\,\alpha \nu \beta} y^\alpha y^\beta
	 + \sum_{s>2} g_2^{1-s/2} \rho^\mu_{s\,\nu}\,,
\end{equation}
where as usual a $0$ denotes quantities associated to the (pseudo-)Riemannian metric $g^0_{\mu \nu}$. Notice that the coefficients in this expansion are typically not traceless even if the $\phi_s$ are. 


\subsection{Sum over spins} 
\label{sub:sum-spins}

Let us take the $\phi_s$ to be small, and work at linear order in them. In this case we can use the formalism developed in Sec.~\ref{sub:def}. Namely, we take $F^2=2g_2= g_{\mu \nu}y^\mu y^\nu$ to be of (pseudo-)Riemannian type, and $\tilde F^2$ to be (\ref{eq:F-y}).

Recall that both $D_\mu$ and $\delta_\mu$ reduce to the usual covariant derivative $\nabla_\mu$ in the (pseudo-)Riemannian case. In particular, these operators have trivial action on the powers of $g_2$ in (\ref{eq:GR-y}). On the other hand, the $y$ derivatives do act nontrivially. 

For $G^\mu$, (\ref{eq:G-def-riem}) gives
\begin{equation}\label{eq:Gs}
	G^\mu_{\mathrm{L}\,s} = \frac12 (\dd \phi^\mu_s - \nabla^\mu \phi_s)+\frac{4-s}4 y^\mu \dd \phi_{s-2}\,. 
\end{equation}
Here and later, the label L stands for ``at linear order in the $\phi_s$''. Recall that $\phi_s=0$ for $s<2$, so the last term in (\ref{eq:Gs}) is actually absent for $G^\mu_3$. 

$G^\mu$ is related to the analogue of a connection, while (\ref{eq:Gs}) is a tensor.\footnote{One of the advantages of working with (\ref{eq:G-def-riem}) is that it makes this property more or less manifest. It is of course possible, but more laborious, to get the same results from (\ref{eq:G}).} The first term in its expansion (\ref{eq:GR-y}) does contain a connection; the following terms are to be thought of as deformations, and it is a familiar result that the difference of two connections is a tensor.

A similar application of (\ref{eq:F-y}) to (\ref{eq:R-def-riem}) gives
\begin{equation}\label{eq:rho-s}
	\rho^\mathrm{L}_s \equiv (\rho^{\mathrm{L}}_{s})^\mu{}_\mu = -\frac12 {\mathcal F}(\phi_s) + \alpha_s \dd^2 \phi_{s-2} \, ,\qquad \alpha_s \equiv \frac18 (s-4)(D+s-4)\,,
\end{equation}
again with L denoting linear order in $\phi_s$. (\ref{eq:rho-s}) are tensors, as manifest from (\ref{eq:R-def}) but not in (\ref{eq:R}). This result was anticipated in the introduction as (\ref{eq:Fgs-d2-intro}).

It is instructive to check invariance under the trivial transformation (\ref{eq:tr-inv}). Since it does not change $F^2$, it should not change the geometric quantities above; let us focus on (\ref{eq:rho-s}). From (\ref{eq:delta-gmu}) (with $\partial_\mu \to \nabla_\mu$) we obtain the identity 
\begin{equation}\label{eq:F(g2gs)}
	{\mathcal F}(-g_2 \phi_{s-2}) = - g_2 {\mathcal F}(\phi_{s-2})+ \left(3-\frac D2 -s \right) \dd^2 \phi_{s-2}\,.
\end{equation}
Fixing a particular $s_0>4$, there are two affected $\phi_s$, which appear in a total of four terms:
\begin{align}
	\delta_\kappa\rho^\mathrm{L} &= g_2^{-s_0/2} \dd^2 \delta_\kappa \phi_{s_0} + g_2^{1-s_0/2}\left(-\frac12 {\mathcal F}(\delta_\kappa\phi_{s_0})+ \alpha_{s_0} \dd^2 \delta_\kappa \phi_{s_0-2}\right) -\frac12 g_2^{2-s_0/2} {\mathcal F}(\delta_\kappa\phi_{s_0-2}) \nonumber\\
	&=g_2^{1-s_0/2} \left(-\alpha_{s_0+2}+ \alpha_{s_0} +\frac12(s_0-3)+\frac D4 \right)\dd^2 \kappa_{s_0-2}\,.
\end{align}
In the second step we have used (\ref{eq:F(g2gs)}). Imposing $\delta_\kappa \rho^\mathrm{L}=0$ for all $s_0$ gives a recurrence relation for the $\alpha_s$, which is indeed satisfied by the value in (\ref{eq:rho-s}). This also shows that the $\dd^2 \phi_{s-2}$ term in (\ref{eq:rho-s}) is unavoidable.\footnote{A curious possibility is to use (\ref{eq:F(g2gs)}) to express $\dd \phi_{s-2}$ in terms of Fronsdal operators, and to substitute in (\ref{eq:rho-s}). This results in $\rho^\mathrm{L}_s = -\frac12 {\mathcal F}(\psi_s)$, where $\psi_s= \frac{\alpha_{s+4}}{\alpha_{s+4}- \alpha_{s+2}} \phi_s - \frac{\alpha_s}{\alpha_{s+2}- \alpha_s} g_2 \phi_{s-2}$, thus getting rid of the second term in (\ref{eq:rho-s}). Inverting this relation one would naively seem to arrive at $F^2= \sum g_2^{1-s/2} \psi_s$. However, the last step only works by summing a series that does not really converge.}

A more complicated computation also gives 
\begin{align}
	\nonumber
	(\rho^{\mathrm{L}}_{s})^\mu{}_\rho = -\frac12 {\mathcal F}^\mu{}_\rho(\phi_s) &+\frac{4-s}4\Big(y^\mu \nabla_\rho \dd \phi_{s-2} + g_{\nu \rho}y^{[\nu} \dd \nabla^{\mu]}\phi_{s-2} -\frac12\delta^\mu_\rho\dd^2  \phi_{s-2} \\
	\label{eq:Rmn-s}
	& -g_{\nu \rho} y^{(\mu} \dd^2\phi^{\nu)}_{s-2}\Big)-\frac1{16}(4-s)(6-s) y^\mu y_\rho \dd^2 \phi_{s-4}\,,
\end{align}
where we introduced a Fronsdal ``precursor''
\begin{equation}\label{eq:fronsdal-mn}
	{\mathcal F}^\mu{}_\rho(\phi_s) = \nabla_\rho \nabla^\mu \phi_s - \nabla_\rho \dd \phi^\mu_s - g^{\mu \nu} \dd \nabla_{[\nu} \phi^s_{\rho]} +\frac12 \dd^2 \phi^\mu_{s\,\rho}\,
\end{equation}
whose trace is ${\mathcal F}(\phi_s)$. Recall from (\ref{eq:R}) that $\rho^\mu{}_\nu$ contains the Finsler analogue of the whole Riemann tensor. We will actually not need it in this paper, but it might be useful in further investigations.

Finally we record the perturbative expression for the mean Cartan (\ref{eq:mean-cartan})
\begin{equation}
 I^{\mathrm{L}\,s}_\mu= \frac14 \phi'_{s\,\mu} + \frac{4-s}8\left((D+s-2) \phi^{s-2}_\mu +y_\mu \phi_{s-2}' \right)+\frac{(4-s)(6-s)}{16} (D+s-4) y_\mu \phi_{s-4}\,
\end{equation}
and for the Landsberg tensor (\ref{eq:landsberg}):
\begin{align}
	L_{\mu \nu \rho}^\mathrm{L\,s} &=\frac14 \dd \phi^s_{\mu \nu \rho} +\frac38 (4-s)\left(g_{(\mu \nu} \dd \phi^{s-2}_{\rho)} + y_{(\mu} \dd \phi^{s-2}_{\nu \rho)}\right)\\
	\nonumber&+ \frac{3(4-s)(6-s)}{16}  \left(y_{(\mu} y_\nu \dd \phi^{s-4}_{\rho)} +y_{(\mu} g_{\nu \rho)}\dd \phi^{s-4} \right) + \frac{(4-s)(6-s)(8-s)}{32} y_\mu y_\nu y_\rho \dd \phi_{s-6}\,.
\end{align}
Again here $I^\mathrm{L}_\nu = \sum_{s>2} g_2^{1-s/2}I^{\mathrm{L}\,s}_\nu$, $L^\mathrm{L}_{\mu\nu \rho} = \sum_{s>2} g_2^{1-s/2}L^{\mathrm{L}\,s}_{\mu \nu \rho}$.



\section{Nonlinearities} 
\label{sec:nonlin}

We now go beyond the linear order in the $\phi_s$. This quickly becomes complicated, but we will at least try to make the structure clear in Sec.~\ref{sub:nonlin}. In Sec.~\ref{sub:spin6} we make the general results more concrete for $\rho_4$ and $\rho_6$.

\subsection{General expansion} 
\label{sub:nonlin}

In this section we raise and lower indices with the unperturbed (pseudo-)Riemannian metric, now denoted by $g^0_{\mu \nu}$ to avoid confusion with the fundamental tensor (\ref{eq:g}) of Finsler geometry. We expand the latter and its inverse as in (\ref{eq:F-y}), (\ref{eq:GR-y}):
\begin{equation}\label{eq:g-y}
	g_{\mu \nu}= g^0_{\mu \nu} + \sum_{s>2} g_2^{1-s/2} q^s_{\mu \nu} \, ,\qquad
	g^{\mu \nu}= g_0^{\mu \nu} + \sum_{s>2} g_2^{1-s/2} p^s_{\mu \nu}\,.
\end{equation}
(\ref{eq:F-y}) gives
\begin{equation}
	q^s_{\mu \nu}= \frac12 \phi^s_{\mu \nu} - \frac{s-4}4 ( \phi_{s-2} g^0_{\mu \nu} + 2 y_{(\mu} \phi^{s-2}_{\nu)} ) + \frac18 (s-4)(s-6) y_\mu y_\nu \phi_{s-4}\,.
\end{equation}
As usual, $\phi_{s<3}$ is understood to be zero; so $q^3_{\mu \nu}= \frac12 \phi^3_{\mu \nu}$, $q^4_{\mu \nu}= \frac12 \phi^4_{\mu \nu}$, 
$q^5_{\mu \nu}= \frac12 \phi^5_{\mu \nu}-\frac14 \phi_3 g^0_{\mu \nu} -\frac12 y_{(\mu} \phi^3_{\nu)}$, in the usual notation (\ref{eq:phis-m}). 
The inverse $g^{\mu \nu}$ can be found by inverting (\ref{eq:g-y}). This is formally similar to several familiar problems in QFT: the answer can be written a bit implicitly as
\begin{equation}\label{eq:pmn}
	p_s^{\mu \nu}= \sum_{\substack{(s_1,\ldots,s_n)\\\text{partitions of }s-2}} (-1)^n (q_{s_1+2}q_{s_2+2}\ldots q_{s_n+2})^{\mu \nu}\,.
\end{equation}
The parenthesis is explicitly $g_0^{\mu\mu_1} q^{s_1+2}_{\mu_1 \nu_1} g_0^{\nu_1 \mu_2}q^{s_2+2}_{\mu_2 \nu_2} g_0^{\nu_2 \mu_3}\ldots q^{s_n+2}_{\mu_n \nu_n}g_0^{\nu_n \nu}$; the partitions are to be considered in all possible orders. The first two (\ref{eq:pmn}) are $p_3^{\mu \nu}= - q_3^{\mu \nu}$, $p_4^{\mu \nu}= -q_4^{\mu \nu} + q^\mu_{3\,\rho} q_3^{\rho \nu}$.

For $G^\mu$, we can work with (\ref{eq:tG}) instead of (\ref{eq:G-def-riem}); it is still more convenient than the original (\ref{eq:G}). The non-linearity in $G^\mu_s$ comes from the inverse $g^{\mu \nu}$ in (\ref{eq:tG}); in terms of (\ref{eq:Gs}) we can write
\begin{equation}\label{eq:G-s-nonlin}
	G_s^\mu = G_{\mathrm{L}\,s}^\mu + \sum_{s'>2} p_{s'}^{\mu \nu} G^{s-s'}_{\mathrm{L}\,\nu}\,,
\end{equation}

The curvature is quite a bit more complicated:
\begin{equation}\label{eq:rho-s-nonlin}
	\rho^\mu_{s\,\rho} = \nabla_\rho G^\mu_s -\frac12 \dd \partial_{y^\rho} G^\mu_s +\frac{s-4}4 y_\rho \dd G^\mu_{s-2}+ Q^\mu_{s\,\rho}\,.
\end{equation}
The term  $Q^\mu_s$, quadratic in $G^\mu_s$, is nastier and reads in general
\begin{align}\label{eq:Q-G}
	Q^\mu_{s\,\rho} = \sum_{s'=3}^{s-1}	&\left(-\frac14 (\partial_{y^\rho}G^\nu)_{s'}(\partial_{y^\nu}G^\mu)_{s-s'+2} +\frac12 G^\nu_{s'} (\partial_{y^\nu}\partial_{y^\rho}G^\mu)_{s-s'+2} \right)\,; \nonumber\\
	(\partial_{y^\rho}G^\mu)_s &= \partial_{y^\rho}G^\mu_s -\frac{s-4}2 y_\rho G^\mu_{s-2} \, ,\\
	(\partial_{y^\nu}\partial_{y^\rho}G^\mu)_s &= \partial_{y^\nu}\partial_{y^\rho} G^\mu_s-\frac{s-4}2 (g_{\nu \rho} G^\mu_{s-2} +2 y_{(\nu}\partial_{y^{\rho)}} G^\mu_{s-2})+\frac14 (s-4)(s-6) y_\nu y_\rho G^\mu_{s-4}\,.	\nonumber
\end{align}
The expression (\ref{eq:rho-s-nonlin}) is only linear and quadratic in $G^\mu_s$, but the $p_s$ appearing in (\ref{eq:G-s-nonlin}) are themselves very non-linear in the $q_s$ and hence in the $\phi_s$. In particular we see that the non-linearity grows with $s$.

\subsection{Spin-six nonlinear equation} 
\label{sub:spin6}

To illustrate the general results above, we now consider the case where only even $s$ are present, and give the first two $\rho_s$.

The first coefficients of the fundamental tensor and its inverse read
\begin{equation}
\begin{split}
	&q^4_{\mu \nu}= \frac12 \phi^4_{\mu \nu} \, ,\qquad q^6_{\mu \nu}= \frac12 \phi^6_{\mu \nu}-\frac12 \phi_4 g^0_{\mu \nu} - y_{(\mu} \phi^4_{\nu)} \,;\\
	&p_4^{\mu \nu}=-\frac12 \phi_4^{\mu \nu} \, ,\qquad 
	p_6^{\mu \nu}= q_4^{\mu \rho}q_{4 \rho}{}^\nu-q_6^{\mu \nu} = \frac14 \phi_4^{\mu \rho} \phi_{4 \rho}{}^\nu +\frac12 g_0^{\mu \nu}\phi_4+ y^{(\mu} \phi_4^{\nu)} -\frac12 \phi_6^{\mu \nu}\,.
\end{split}
\end{equation}
The first spray coefficients are
\begin{equation}
	G_4^\mu  = \frac12(\dd \phi^\mu_4- \nabla^\mu \phi_4) \, ,\qquad
	G_6^\mu = \frac12(\dd \phi^\mu_6- \nabla^\mu \phi_6)-\frac12y^\mu \dd \phi_4 -\frac12g^{\mu \nu}_4 G_{4\,\nu}\,.
\end{equation}
In particular, the $s=4$ result coincides with the linear result (\ref{eq:Gs}).

As for the curvature, again the $s=4$ result coincides with the linear (\ref{eq:Rmn-s}), (\ref{eq:fronsdal-mn}), while for $s=6$ a quadratic term appears:
\begin{align}
	\rho^\mu_{4\,\rho} &= \nabla_\rho G^\mu_4 -\frac12 \dd \partial_{y^\rho}G^\mu_4= (\rho^{\mathrm{L}}_4)^\mu{}_\rho\, ,\\
	\nonumber
		\rho^\mu_{6\,\rho} &= \nabla_\rho G^\mu_6 -\frac12 \dd \partial_{y^\rho} G^\mu_6 +\frac12 y_\rho \dd G^\mu_4
		-\frac14 \partial_{y^\rho}G^\nu_4 \partial_{y^\nu}G^\mu_4 +\frac12G^\nu_4 \partial_{y^\nu}\partial_{y^\rho} G^\mu_4 =
		(\rho^{\mathrm{L}}_6)^\mu{}_\rho + {\mathcal Q}^\mu{}_\rho (\phi_4)\,,
\end{align}
where
\begin{equation}
	 {\mathcal Q}^\mu{}_\rho (\phi_4)= -\frac12 \nabla_\rho(\phi_4^{\mu \nu}G^4_\nu) +\frac14 \dd (\phi_4^{\mu \nu}{}_\rho G^4_\nu + \phi_4^{\mu \nu}\partial_{y^\rho}G^4_\nu) -\frac14 \partial_{y^\rho}G^\nu_4 \partial_{y^\nu}G^\mu_4
	 + \frac12 G^\nu_4 \partial_{y^\nu}\partial_{y^\rho}G^\mu_4\,.
\end{equation}
This ${\mathcal Q}$ now denotes the part that is quadratic in $\phi_4$ rather than in $G^\mu_4$, so it contains more terms than the $Q$ in (\ref{eq:Q-G}).

Notice that the $\rho^\mu_{s\, \rho}$ are not symmetric if we lower the first index with $g^0_{\mu \nu}$ (as we are doing in this section). Indeed the symmetric property advertised in \ref{sub:conn} holds if we lower the index with the full fundamental Finsler tensor, $\rho_{\mu \rho}\equiv g_{\mu \nu}\rho^\nu{}_\rho$. In our spin expansion, the $s=4$ term of this object reads $g^0_{\mu \nu}\rho^\nu_{4\,\rho}+\frac12 \phi^4_{\mu \nu}R^\nu{}_{\alpha \rho \beta}y^\alpha y^\beta$; this can indeed be shown to be symmetric using (\ref{eq:[nn]}).\footnote{For the much more laborious $s=6$ check, I used the software package xAct.}

Finally, taking the trace:
\begin{equation}
	\rho_4 =-\frac12 {\mathcal F}(\phi_4) \, ,\qquad \rho_6 = -\frac12 {\mathcal F}(\phi_6) + \frac14(D+2) \dd^2 \phi_4+ {\mathcal Q}(\phi_4)\,.
\end{equation}
Formally this looks like a cubic interaction among two $s=4$ and one $s=6$ fields. 



\section{Challenges for Finsler dynamics} 
\label{sec:challenges}

Given the appearance of the Fronsdal operator in Sec.~\ref{sec:ff}, we now investigate whether a Finsler action can indeed propagate degrees of freedom with higher spins. Unfortunately this will give mixed results.

As commented in the introduction, there are strong arguments against a flat space local covariant theory of massless higher spins. So the challenges we will find are not unexpected. Nevertheless, we will start from flat space for simplicity. We will find later that the situation does not seem to improve much for AdS.

\subsection{Equations of motion} 
\label{sub:eom}

For pure GR, Einstein's equations set to zero the Ricci tensor. This suggests to take for pure Finsler gravity 
\begin{equation}\label{eq:rutz}
	\rho=0\,,
\end{equation}
sometimes called Rutz equation \cite{rutz}. For us this has the advantage that we know it already to be related to the Fronsdal operator. A natural variant is $\rho - k F^2=0$, for $k$ a constant, which is similar to pure GR with a cosmological constant.

On the other hand, it was argued in \cite{hohmann-pfeifer-voicu,hohmann-pfeifer-voicu2} that this cannot be obtained from an action. Another natural possibility is to write the closest possible analogue of the Einstein--Hilbert action for pure GR. $\rho$ also contain $y$ variables, and to obtain an analogue of the Ricci scalar we would like to somehow trace over them. This can be achieved by integrating over them, leading to 
\begin{equation}\label{eq:S-pw}
	S= \int \dd^D x\,\dd^{D-1}\hat y\, g \,\frac{\rho}{F^2} \,.
\end{equation}
(See \cite[11.3.1]{shen-shen} and \cite{pfeifer-wohlfarth}.)
The measure is the product of the usual $\sqrt{g}\dd^D x$ and of $\sqrt{g}\dd^{D-1}\hat y$ along the fiber of the sphere tangent bundle $SM$. The $F^2$ is included to make the integrand of degree zero, so that it is defined on the fiber of that bundle.  The variation of (\ref{eq:S-pw}) gives
\begin{equation}\label{eq:pw}
	 \rho = \frac1{D+2}F^2 g^{\mu \nu}\left(\partial_{y^\mu} \partial_{y^\nu}  \rho + 2 D_\mu \dd I_\nu -2 \dd I_\mu \dd I_\nu +2 \partial_{y^\mu}  \dd^2 I_\nu\right)\,.
\end{equation}
We will see that (\ref{eq:rutz}) and (\ref{eq:pw}) can be analyzed along similar lines.\footnote{A third possibility was proposed in \cite{garciaparrado-minguzzi}, which is however chosen in such a way that its classical solutions coincide with those of ordinary GR. Thus we will not consider it here.}

In the rest of this section we will analyze the classical solutions of this theory, perturbatively around Minkowski space. In the similar problem for GR, the equations of motion can be approximated by taking (\ref{eq:Ricci0-def}) for $g_{\mu \nu}= \eta_{\mu \nu}$. Crucially, the infinitesimal diffeomorphisms (\ref{eq:deltaxi-h}) can be used to set $\partial^\nu h_{\mu \nu}-\frac12 \partial_\mu h=0$, after which the equation of motion reads simply
\begin{equation}\label{eq:de2h=0}
	\partial^2 h_{\mu \nu}=0\,.
\end{equation}
So there are no massive excitations. Diffeomorphisms can be further used to restrict the massless ones to $\partial^\mu h_{\mu \nu}=u^\mu h_{\mu \nu}=0$, with $u$ a timelike vector. These results avoid the presence of modes with the wrong sign in the action, which would be problematic for quantization. (While pure GR is ultimately not sensible quantum mechanically because of its non-renormalizability, it can at least be used as a good effective quantum field theory.)

We thus expect the role of gauge transformations to be crucial. The presence of the Fronsdal operator ${\mathcal F}$ in (\ref{eq:R-def-riem}), (\ref{eq:rho-s}) seems promising in that respect: ${\mathcal F}$ does have an infinite-dimensional kernel, or in other words it has gauge transformations. In the next subsection we examine whether these can be considered gauge transformations for our Finsler equations of motion.


\subsection{Broken Fronsdal gauge transformations} 
\label{sub:gauge}

The appearance of the Fronsdal operator in (\ref{eq:rho-s}) might give hope that gauge transformations of the form $\delta \phi_s = \dd \lambda_{s-1}$ can be introduced in Finsler geometry, at least for a small perturbation around flat space $F^2=2 \eta_2$ (where (\ref{eq:delta-Fs}) is valid). It is not clear why the $F^2$ parameterization (\ref{eq:F-y}) should be invariant under such a transformation.\footnote{\cite[(5.9)]{dewit-freedman} notes that the invariance of proper time $\int \dd \tau F$ can be approximately restored by simultaneously changing the position $x^\mu(\tau)$ in a velocity-dependent fashion.} In other words, the gauge transformations of Finsler geometry are only diffeomorphisms. A priori it might be possible, however, that $F^2$ changes in a way that does not modify the geometry in a physically relevant way. We would then declare such transformations to be additional gauge transformations.
This might perhaps be similar to the notion of projective equivalence mentioned in Sec.~\ref{sub:def}.\footnote{The infinitesimal counterpart of (\ref{eq:tG}) is $\delta G^\mu = \delta P y^\mu$ with $\delta P= (2g_2)^{-1}\dd \delta F = (2 g_2^3)^{-1/2} \dd \delta F^2$. From (\ref{eq:R-def0}) we now find $\delta \rho= (1-D/2) \dd \delta P$; so $\rho$ is not invariant under a projective equivalence.} 

However, the term $\dd^2 \phi_{s-2}$ in (\ref{eq:rho-s}) should give one pause. Including the Fronsdal term, the total transformation for a Finsler deformation of Minkowski space reads 
\begin{equation}\label{eq:delta-rhos-l}
	\delta \rho^\mathrm{L}_s = \dd^3 \left(-\frac14 \lambda'_{s-1}+ \alpha_s \lambda_{s-3}\right) \,.
\end{equation}
One might try to cancel the two terms in the parenthesis against each other. The structure looks similar to that of a St\"uckelberg mechanism, where a mass term locks together the gauge transformations of two fields of different spin. It is also reminiscent of the compensator fields in \cite{francia-sagnotti,francia-sagnotti2,demedeiros-hull,sagnotti-tsulaia}, where an equation of the type ${\mathcal F}(\phi_s)= \dd^3 {\mathcal H}_s$ appears, with $\delta {\mathcal H}_s= \lambda'_{s-1}$.\footnote{In the AdS case an extra term is present because of (\ref{eq:deltaF-AdS}); we will deal with this at the end. For theories of Vasiliev type, a certain boundary condition for the scalar can lead to a Higgs mechanism whose Goldstone field is a bound state of higher spin fields \cite{girardello-porrati-zaffaroni}.}

Unfortunately, this combined gauge transformation is nothing but a particular case of the trivial redundancy (\ref{eq:tr-inv}), which is only present when the $\phi_s$ are not traceless. This is easiest to see when we only transform two spins, $\delta \phi_{s_0}= \dd \lambda_{s_0-1}$, $\delta \phi_{s_0-2}=\dd\lambda_{s_0-3}$, and the two only contain the same irreducible spin-$(s_0-3)$ representation: $\lambda^0_{s_0-1}=0$, $\lambda''_{s_0-1}=0$, and $\lambda'_{s_0-3}=0$. Setting $\delta \rho^\mathrm{L}$ to zero gives $\lambda_{s_0-3}= \lambda'_{s_0-1}/(6-D-2s)$, which upon taking $\dd$ reduces to (\ref{eq:tr-inv}). 

A faster way to proceed is to assemble all the $\lambda_s$ into a single 
\begin{equation}\label{eq:total-lambda}
	\Lambda= \sum_{s>1} g_2^{1-s/2} \lambda_{s-1}\,,
\end{equation}
so that now
\begin{equation}
	\delta F^2 = \dd \Lambda\,.
\end{equation}
Notice that $\Lambda$ is homogeneous in $y$ of degree one, since $F^2$ is of degree two.
 When we perturb around flat (pseudo-)Riemannian space, $F^2=2 \eta_2$, and use (\ref{eq:R-def-riem}), the same steps as in (\ref{eq:delta-Fs}) give
\begin{equation}\label{eq:delta-rho-l}
	\delta \rho = -\frac14 \dd^3 \Lambda' \, ,\qquad \Lambda'= \eta^{\mu \nu}\partial_{y^\mu} \partial_{y^\nu} \Lambda\,. 
\end{equation}
The strategy above now corresponds to demanding $\Lambda'=0$. This is not simply the sum over the individual $\lambda'_{s-1}$, because the factors $g_2^{1-s/2}$ in (\ref{eq:total-lambda}) also give a contribution when we take the derivatives with respect to $y^\mu$. (This is the subtlety alluded to above (\ref{eq:R-def-riem}).) Rather:
\begin{equation}\label{eq:l'}
	\Lambda'= \sum_{s>1}g_2^{1-s/2} \left(\lambda'_{s-1} - 4 \alpha_s \lambda_{s-3}\right)\,.
\end{equation}
The coefficients in this expansion are not necessarily traceless, so to set $\Lambda'=0$ we need to use the logic leading to (\ref{eq:exp-proj}). This gives a relation $\lambda^0_{s-3}+ \lambda^1_{s-1}+\ldots=0$, which upon using (\ref{eq:exp-proj}) gives $\Lambda=0$. Alternatively, the discussion in Sec.~\ref{sub:param} tells us that the parameterization (\ref{eq:total-lambda}) is redundant unless we impose that the $\lambda_{s-1}$ are irreducible:
\begin{equation}\label{eq:l's=0}
	\lambda'_{s-1}=0\,.
\end{equation}
In other words, this choice can be made without loss of generality. Notice that this is the infamous traceless condition on the Fronsdal gauge transformations, which here appears rather naturally. In view of (\ref{eq:l'}), $\Lambda'=0$ implies $\lambda_{s-1}=0$ for all $s$. 

An even nicer way of seeing the failure of this strategy is to notice that $\Lambda'= \eta^{\mu \nu}\partial_{y^\mu} \partial_{y^\nu} \Lambda= 0$ is demanding $\Lambda$ to be a harmonic function of the $y$, while below (\ref{eq:total-lambda}) we observed it to be homogeneous of degree one. These two demands are only met by a linear function $\Lambda= \lambda_1$, which corresponds to $\delta g_2 = \dd \lambda_1$, the familiar diffeomorphisms. Ignoring this possibility as we did above, we recover that $\Lambda'=0$ implies $\Lambda=0$.

We have concluded that we cannot set to zero the parenthesis in (\ref{eq:delta-rhos-l}), or $\Lambda'$ in (\ref{eq:delta-rho-l}). In other words, we have tried to find gauge transformations with an algebraic constraint (similar to the traceless constraint for Fronsdal), and we failed. We could still hope that the full (\ref{eq:delta-rhos-l}) can somehow be made zero, by playing with the differential operators. However, $\dd^3$ has a finite-dimensional kernel, consisting of quadratic polynomials in the $x^\mu$. Thus we also cannot find gauge transformations this way.\footnote{We can apply a similar argument to projective equivalence to flat space. The variation of (\ref{eq:G-def-riem}) gives $\delta G_\mathrm{L}^\mu = \frac12 \dd^2 \partial_{y_\mu}\Lambda$. From below (\ref{eq:tG}) we find  $\delta G^\mu= P y^\mu$. If we ignore the finite-dimensional kernel of $\dd^2$, we are led to postulate $\partial_{y_\mu}\Lambda= y^\mu Q$, with $\dd^2 Q= 2P$. This implies a ``purely radial'' dependence, $\Lambda=\Lambda(\eta_2)$, and the homogeneity property lets us conclude $\Lambda= \lambda_0\sqrt{\eta_2}$, which would correspond to the gauge transformations of a spin-1 field, had we included it in (\ref{eq:F-y}). This is compatible with \cite[Prop.~3.4.8]{chern-shen}; notice however that many other Finsler metrics projectively equivalent to flat space are given in that book, not obtained by gauge transformations of Fronsdal type.}

We now also analyze this possibility using the parameterization in terms of sum over spins. It is lengthier, but instructive, and also a good warm-up for the next subsection. We earlier concluded that we can always fix $\lambda'_{s-1}=0$. Moreover we recall from Sec.~\ref{sub:param} that $\phi_s'=0$ can similarly be assumed without loss of generality. Since $(\dd \lambda)'= \dd \lambda'+2 \partial_\mu \lambda^\mu$, we also conclude (as in \cite{skvortsov-vasiliev}) that
\begin{equation}\label{eq:trls=0}
	\partial_\mu \lambda^\mu_{s-1}=0\,.
\end{equation}
Setting $\delta \rho^\mathrm{L}=0$ does not necessarily imply that the coefficients $\delta\rho^\mathrm{L}_s$ of its $g_2$ expansion vanish. 
The latter are given by (\ref{eq:delta-rhos-l}) and are not necessarily traceless; thus their trace parts can mix in the total sum $\sum_{s>2} g_2^{1-s/2} \delta\rho^\mathrm{L}_s$. As discussed near (\ref{eq:exp-proj}), the series can be equivalently rewritten with coefficients $\sum_{j=0}^\infty \delta\rho^\mathrm{L}_{s+2j,j}$ (recalling the notation (\ref{eq:phi-s0})), which are now automatically traceless and can be set to zero. Given (\ref{eq:l's=0}), (\ref{eq:trls=0}) and recalling that we are taking $g_2=\eta_2$, we compute $(\dd^3 \lambda)'=6\dd \partial^2 \lambda$, $(\dd^3 \lambda)''=0$. So gauge invariance of $\rho^\mathrm{L}$ requires
\begin{equation}\label{eq:gauge?}
	\left[\alpha_s \dd^3\lambda_{s-3} -\frac{6\alpha_{s+2}}{D+2s} \dd \partial^2 \lambda_{s-1} \right]_0 =0\,.
\end{equation}
One might hope that the two terms can sum to zero; this would lead to a recursive law on the $\lambda_{s-1}$. We now show that this cannot happen. The action of $\dd$ can be diagonalized after going to Fourier transform. Consider first $p^2< 0$. We can take $p^\mu=(p^0,\,0,\,\ldots,\,0)$, so
\begin{equation}\label{eq:dp0}
	\dd = -\ii y^0 p_0 \, ,\qquad \dd^\dagger = \ii p_0 \partial_{y^0}\,. 
\end{equation}
(\ref{eq:trls=0}) gives us $\partial_{y^0} \lambda_{s-1}=0$. The traceless projection in (\ref{eq:gauge?}) adds a slight complication, which is reviewed in App.~\ref{app:d}. The two terms in (\ref{eq:gauge?}) can be viewed (forgetting the constants) as $(z\dd)^3 \lambda_{s-3}$ and $(z\dd) \lambda_{s-1}$, with $z \lambda \equiv \lambda_0$ the traceless projector. On our $y^0$-independent $\lambda$'s, the action of $(z\dd)^j$ is proportional to multiplication by $y_0^j$. This means that the two terms in (\ref{eq:gauge?}) have $y^0$ dependence proportional to $y_0^3$ and $y_0$; thus the two cannot be equal, unless they are both zero. In the case $p^2=0$, the second term in (\ref{eq:gauge?}) is directly zero. Moreover, $z\dd^3= (z\dd)^3$ is invertible, as also shown in App.~\ref{app:d}. Thus there are no solutions to (\ref{eq:gauge?}).

We can also try a similar analysis on a maximally symmetric space such AdS. This a priori might look more promising, in light of the aforementioned no-go arguments. Here is what changes. First, covariant derivatives don't commute, and that complicates the trace $(\dd^3 \lambda)'$, which is now $(6\dd(2 \nabla^2-\Delta)-2[\dd,\nabla^2]) \lambda_{s-1}$. Second, now $\delta {\mathcal F}\neq 0$ even if $\lambda'_{s-1}=0$, as we saw back in (\ref{eq:deltaF-AdS}). Third, it makes sense to consider now the variant of (\ref{eq:rutz}) discussed below it, $\rho-k F^2=0$. This however can be reabsorbed in (\ref{eq:deltaF-AdS}) by redefining $\Lambda$. All in all (\ref{eq:gauge?}) is changed to
\begin{equation}
	\left[-\frac12\Lambda \mu_s^2 \dd \lambda_{s-1} + \alpha_s \dd^3 \lambda_{s-3} -\frac{\alpha_{s+2}} {D+2s}\dd( 6 \nabla^2 + \gamma_s)\lambda_{s-1} \right]_0 =0 \,.
\end{equation}
where $\gamma_s= \Lambda(6 (s-1)(D+s-3)+ D+2s-3))/(D-1)$. Now we cannot perform a Fourier transform to analyze the resulting condition; we should work in a suitable eigenbasis for AdS. I have not carried out this analysis, but the problem looks similar enough to (\ref{eq:gauge?}) that at small distances (where Fourier transform can be used at least approximately) it should reduce to the one for flat space. 

In this subsection we have concluded that in spite of (\ref{eq:R-def-riem}), (\ref{eq:rho-s}), the Fronsdal gauge transformations $\delta \phi_s = \dd \lambda_{s-1}$ don't survive in Finsler geometry except as the trivial redundancies (\ref{eq:tr-inv}). It is still possible that there is some other gauge transformation, perhaps obtained by complementing the Fronsdal ones by additional terms involving $\dd^\dagger$ or other operators. In any case, in the next subsection we will sketch a linearized analysis of the Finsler equations of motion.


\subsection{Perturbations} 
\label{sub:per}

Let us first consider (\ref{eq:rutz}).  We use the power series parameterization (\ref{eq:F-y}), taking $\phi_s'=0$. We found the linearized $\rho^\mathrm{L}$ in (\ref{eq:rho-s}). Just like for the earlier discussion of gauge transformations, $\rho^\mathrm{L}=0$ does not necessarily imply that the $\rho^\mathrm{L}_s$ vanish. 
As discussed near (\ref{eq:exp-proj}), the series can be equivalently rewritten with coefficients $\sum_{j=0}^\infty \rho^\mathrm{L}_{s+2j,j}$, which are now automatically traceless and can be set to zero. Recalling again (\ref{eq:phi-dec}), we use (\ref{eq:d-comm}) to compute
\begin{equation}
	(\dd^2 \phi_s)'' = -4 ({\mathcal F}\phi_s)'= 8 (\dd^\dagger)^2 \phi_s  \, ,\qquad ({\mathcal F}\phi_s)''=0\,.  
\end{equation}
So $\rho^\mathrm{L}_s$ only has single and double traces, and $\rho^\mathrm{L}_{s+2j,j}=0$ for $j\geqslant 3$. This leads to the linearized equation of motion
\begin{equation}\label{eq:lin-eom-s}
	\left[-\frac12 {\mathcal F}\phi_s +\alpha_s\dd^2 \phi_{s-2}
	+ f^3_s ({\mathcal F}\phi_{s+2})' + f^4_s (\dd^2 \phi_s)'
	\right]_0=0\,,
\end{equation}
where $f^3_s = (2(D+2s))^{-1}(1-4 \alpha_{s+4}/(D+2s+2))$, $f^4_s =- \alpha_{s+2}/(D+2s)$.

We now turn to (\ref{eq:pw}). The linearization of the second term is simply $\rho'$. Among the terms involving $I_\mu$, we can ignore the quadratic one; the others give
\begin{align}
	\nonumber
	[g^{\mu \nu}(D_\mu \dd I_\nu &+ \partial_{y^\mu}  \dd^2 I_\nu)]^\mathrm{L}= (\nabla_\mu \dd \partial_{y_\mu} + \partial_{y_\mu} \dd^2 \partial_{y_\mu} ) t \dd F^2 = (2 \nabla_\mu \dd \partial_{y_\mu} + \dd \dd^\dagger + \dd^2 t)t \delta F^2 = \\ 
	&=\left({\mathcal F}-\frac12 t \dd^2\right)t \delta F^2\,.
\end{align}
We used $1/2[t,\dd^2]=\nabla^2 + \nabla_\mu \dd \partial_{y_\mu}+ \dd \dd^\dagger$, which in turn follows again from (\ref{eq:d-comm}).
Overall the linearized version of (\ref{eq:pw}) reads
\begin{equation}
	\left({\mathcal F} -\frac2{D+2} g_2 (t {\mathcal F} -4 {\mathcal F}t +2 t \dd^2 t ) \right) \delta F^2=0\,.
\end{equation}
A lengthier computation gives an equation again of the form (\ref{eq:lin-eom-s}), with more complicated coefficients. 

Given this qualitative similarity, from now on we will focus on (\ref{eq:lin-eom-s}). We take the pseudo-Riemannian geometry to be flat, $g_2=\eta_2$. This simplifies $(\dd^2 \phi_s)'= (2 \partial^2 + \dd \dd^\dagger)\phi_s$. Schematically the equation is now of the form $[(\partial^2+ \dd \dd^\dagger)\phi_s + \dd^2 \phi_{s-2}+ (\dd^\dagger)^2 \phi_{s+2}]_0=0$. This can be clarified further by going to momentum space as in the previous subsection. The simplest case is $p^2< 0$, corresponding to massive modes. In GR these are gauged away, as reviewed near (\ref{eq:de2h=0}), but for Finsler we haven't found a gauge transformation that can achieve a similar result. We can use (\ref{eq:dp0}) and the explicit decomposition in App.~\ref{app:d}, which tackles the traceless projection in (\ref{eq:lin-eom-s}). Each $\phi_s$ is decomposed under the spatial $\mathrm{SO}(D-1)$ as a sum $\sum_k Y_{sk} \hat \phi^0_{s,k}$, where $\hat\phi^0_{s,k}$ is traceless, degree $k$ and $y^0$-independent, while $Y_{sk}$ is a polynomial whose highest-order $y_0$ term is $y_0^{s-k}$. The operators $z\dd$ and $z\dd^\dagger$ (where $z$ is the traceless projector, $z \phi_s= \phi_{s,0}$) are proportional to the identity on each $Y_{sk} \hat \phi^0_{s,k}$. 

For concreteness let us look at the lowest degrees in (\ref{eq:lin-eom-s}), keeping only even $s$. For $s=2$, the equation reads $[(\dd^\dagger)^2 \phi_4]_0=0$. The $\hat \phi_{4,4}$ and $\hat \phi_{4,3}$ don't appear, while $\hat \phi_{4,k\leqslant 2}=0$. For $s=4$, schematically we have $[(\dd^\dagger)^2 \phi_6+ (\partial^2 + \dd \dd^\dagger )\phi_4]_0=0$. Now $\hat \phi_{6,6}$ and $\hat \phi_{6,5}$ don't appear, while for $k\leqslant 4$, $\hat\phi_{6,k}\propto(\partial^2+c_{6,k})\phi_{4,k} $, for $c_{6,k}$ some constants; in particular, $\hat \phi_{6,k\leqslant 2}=0$. Continuing in this fashion we see that $\hat\phi_{s,k\leqslant 2}=0$, while the $\hat\phi_{s,s}$ and $\hat\phi_{s,s-1}$ are free, and determine the $\hat \phi_{s,k}$ with $3\leqslant k \leqslant s-2$. 

Massive modes would then exist for any mass, and they would depend on many free variables. In particular there would be many $\phi_s$ not transverse to $p^\mu$. This would create issues when trying to quantize this theory. The analysis for massless modes is more complicated, but gives a similar result. 

The result of this perturbative analysis is disappointing, but perhaps also expected. Without a gauge transformation, the Finsler equations of motion would be problematic from a perturbative QFT standpoint.


\subsection{Discussion} 
\label{sub:disc}

The appearance of Fronsdal operators in the Finsler Ricci curvature is encouraging. But the associated higher spin gauge transformations are broken.\footnote{Holography indicates that higher spin gauge transformations are exactly preserved only in the Vasiliev theory for a particular choice of boundary conditions \cite{maldacena-zhiboedov}; even their weak breaking at the quantum level is quite constrained \cite{maldacena-zhiboedov2,giombi-minwalla-prakash-trivedi-wadia-yin,aharony-gurari-yacoby,girardello-porrati-zaffaroni}.} Finsler geometry only has diffeomorphisms as gauge transformations. 

A pessimistic conclusion would be that the Fronsdal operator appears only for purely mathematical reasons, driven by its similarity with the spin-two situation. 

More optimistically, perhaps some sort of gauge transformation can still be found. While the most natural guess for gauge transformations was excluded in Sec.~\ref{sub:gauge}, I am not excluding that more exotic possibilities might work. Actually the presence of many free parameters in the previous subsection seems to indicate that some sort of gauge transformation should exist. It should be possible to find operators that have those free paramters as their image. We would also like, however, for such a gauge transformation to have some sort of geometrical interpretation. It would be difficult otherwise to imagine that it can survive beyond the linear order. After all, the possibility of some natural interacting structure is the most interesting possible advantage of using Finsler geometry for higher spins, as illustrated in Sec.~\ref{sec:nonlin}.

While the Finsler Ricci $\rho$ seems important in view of its connection to the Fronsdal operator, another possibility is that it might have to be modified by additional terms. Even more radically, one might have to modify the central tenets of Finsler geometry altogether, as advocated in \cite{hull-W}. In a similar spirit, one might try to adapt the Hamiltonian approach in \cite{segal,ponomarev-symplectic}. Its advantage is that higher-spin symmetries are manifest as canonical transformations. It is possible to define a so-called ``coFinsler'' distance in this way, and it would be interesting to see how the present results are modified in that language. One should however recall that the results in \cite{segal,ponomarev-symplectic} were partially negative (see also \cite{ivanovskiy-ponomarev,ivanovskiy-ponomarev2}). 

It might be helpful to try and recast in terms of Finsler geometry the existing higher spin theories such as those by Vasiliev \cite{fradkin-vasiliev1,fradkin-vasiliev2,vasiliev1,vasiliev2,vasiliev3}, the simpler alternatives in three dimensions \cite{campoleoni-fredenhagen-pfennninger-theisen,campoleoni-fredenhagen-pfennninger-theisen2}, or the more recent self-dual theories \cite{ponomarev-skvortsov,ponomarev-selfdual,skvortsov-tran-tsulaia,krasnov-skvortsov-tran}.\footnote{In a twistor-inspired approach to gravity, Finsler geometry appears rather naturally \cite{mason-bach}.}
Finally, the higher spin symmetries might be realized around a solution where all the higher spin fields are non-zero, rather than around a (pseudo-)Riemannian geometry as attempted here.\footnote{We thank M.~Montero and C.~Pfeifer for this suggestion.}

In any case, after finding a geometrical quantity that is invariant under a set of infinite transformations that extends diffeomorphism invariance, we would declare such transformations to be additional gauge transformations, identifying different Finsler geometries that are equivalent as far as physics is concerned. Time will tell whether any of this can be achieved.



\section*{Acknowledgements}

It's a pleasure to thank J.~Calder\'on Infante, S.~Giombi, C.~Hull, L.~Mason, E.~Minguzzi, M.~Montero, C.~Pfeifer, D.~Ponomarev, A.~Sagnotti, A.~Zaffaroni for very useful discussions and correspondence. My interest in Finsler geometry was rekindled by an ongoing collaboration with B.~De Luca, N.~De Ponti, A.~Mondino, to whom I am very grateful. I would also like to thank the participants in the 2024 Cook's Branch Workshop, to whom a preliminary version of these findings was presented. I am supported in part by INFN and by MIUR-PRIN contract 2022YZ5BA2.

\appendix

\section{More details on the symmetrized derivative} 
\label{app:d}

In this appendix we give a more explicit discussion of some properties of the operator $\dd= y^\mu \nabla_\mu$.

We introduce
\begin{equation}
	 z \phi = (\phi)_0\,,
\end{equation}
where the ${}_0$ denotes the traceless part as in (\ref{eq:phi-s0}). We now act with $\dd$ on the decomposition (\ref{eq:phi-dec}), and then we take the traceless part of the result. Since $\phi'_{s,k}=0$, we have $(\dd \phi_{s,k})'=2 \dd^\dagger \phi_{s,k}$, $(\dd \phi_{s,k})''=0$. Then
\begin{equation}
	(\dd \phi_s)_0 = (\dd \phi_{s,0})_0 = \dd \phi_{s,0} + 2 t_{s+1,1} g_2 \dd^\dagger \phi_{s,0}\,.
\end{equation}
We can also rewrite this as
\begin{equation}
	[z,\dd] \phi_s = 2 t_{s+1,1} g_2 \dd^\dagger z \phi_s\,. 
\end{equation}
In particular it follows
\begin{equation}
	[z,\dd]z \phi_s =0\,,
\end{equation}
which in turn implies
\begin{equation}\label{eq:zdk}
	(z \dd)^k = z \dd^k \, ,\qquad z\dd z \dd^\dagger = z \dd \dd^\dagger\,.
\end{equation}

We now specialize to flat space. Working in Fourier transform, there are two cases to consider: $p^\mu$ massive or massless. In the former case, by a Lorentz transformation we can always take the momentum to be purely along time, $p^0\neq 0$, $p^i=0$, $i=1,\,\ldots,\,D-1$. Then $\dd= -\ii y^0 p_0$, $\dd^\dagger = \ii p_0 \partial_{y^0}$. A traceless $\phi_{s,0}$ can be decomposed in $\mathrm{SO}(d-1)$ representations as
\begin{equation}\label{eq:phi-Y}
	\phi_{s,0}= \sum_k Y_{sk} \hat \phi^0_{s,k}(\vec y) \, ,\qquad
	Y_{sk}= y_0^{s-k} + a_{1sk} \delta_2 y_0^{s-k-2}+\ldots
\end{equation}
where $\hat\phi^0_{s,k}$ is traceless and degree $k$; the $s$ is kept only as a label. The hat denotes dependence on the $y^i$ alone, $\partial_{y^0}\hat \phi_{s,k}=0$, and the coefficients are explicitly
\begin{equation}
	a_{1sk}=\frac{(s-k)(s-k-1)}{D+2k-1} \, ,\qquad
	a_{jsk}=\frac{(s-k-2j+2)(s-k-2j+1)}{j(D+2j+2k-3)} a_{j-1,sk}\,.
\end{equation}
Here $\delta_2=\frac12 \delta_{ij} y^i y^j$, and the coefficients are determined by $(\phi_{s,0})'=0$. (An expansion similar to (\ref{eq:phi-Y}) can be found in \cite{fronsdal}.) One finds 
\begin{equation}\label{eq:dYp}
	(\dd Y_{sk} \hat \phi^0_{sk})_0 = -\ii p_0 \frac{D+s+k-2}{D+2s-2} Y_{s+1,k} \hat \phi^0_{s,k}\,
	\, ,\qquad(\dd^\dagger Y_{sk} \hat \phi^0_{sk})_0 =\ii p_0 (s-k)Y_{s+1,k} \hat \phi^0_{s,k}\,,
\end{equation}
reproducing for $D=4$ formulas in \cite{fronsdal}. Using (\ref{eq:zdk}) one can also find similar expressions for $(\dd^k Y_{sk} \hat \phi^0_{sk})_0= (z\dd)^k \hat \phi^0_{sk}$. 

For massless momentum we can take $p^-=0=p^i$; now $\dd = -\ii y^- p_-$. (\ref{eq:phi-Y}) is replaced by a decomposition in terms of $\mathrm{SO}(d-2)$ representations:
\begin{equation}\label{eq:phi-Z}
	\phi_{s,0}= \sum_k Z_{j_+sk} \hat \phi^0_{s,k}(\vec y) \, ,\qquad
	Z_{j_+sk}= y_+^{j_+}y_-^{s-k-j_+} + b_{1sk} \delta_2 y_+^{j_+-1}y_-^{s-k-j_+-1}+\ldots\,,
\end{equation}
where now $\delta_2$ is the transverse $(D-2)$-dimensional metric, and the coefficients are given by
\begin{equation}
	b_{1sk}= \frac{j_+(s-k-j_+)}{D+2k-2} \, ,\qquad b_{jsk}= \frac{(j_+-j)(s-k-j_+-j)}{j(D+2j-4+2k)}b_{j-1,sk}\,.
\end{equation}
(\ref{eq:dYp}) is replaced by
\begin{equation}\label{eq:dZp}
	(\dd Z_{j_+sk} \hat \phi^0_{sk})_0 = 2\ii p_- \frac{D+2k+2j_+}{D+2s} Z_{j_++1,s+1,k} \hat \phi^0_{s,k}\,.
\end{equation}

From (\ref{eq:dYp}), (\ref{eq:dZp}) it now follows that $\dd$ has no kernel. 


\bibliography{at}
\bibliographystyle{at}

\end{document}